\documentclass[prb,amsmath,showpacs,twocolumn]{revtex4-1}
\usepackage{amsmath, amssymb, graphicx, bm}
\usepackage[usenames,dvipsnames,svgnames,table]{xcolor}
\usepackage[colorlinks=true,linkcolor=RoyalBlue,citecolor=RoyalBlue]{hyperref}
\usepackage{bbold}
\usepackage{mathtools}
\usepackage{relsize}

\def\bs{\boldsymbol}

\def\mr{\mathrm}
\def\mc{\mathcal}
\def\mf{\mathfrak}
\def\pmm{\scalebox{0.5}[0.5]{\( \pm \)}}
\def\mperp{\scalebox{0.5}[0.5]{\( \perp \)}}
\def\sm{\scalebox{0.5}[0.5]{\( - \)}}
\def\smm{\scalebox{0.5}[0.5]{\( + \)}}

\begin{document}

\title{Spin transfer and spin pumping in disordered normal metal-antiferromagnetic insulator systems}

\author{Sverre A. Gulbrandsen }
\affiliation{Center for Quantum Spintronics, Department of Physics, Norwegian University of Science and Technology, NO-7491 Trondheim, Norway}

\author{Arne Brataas}
\affiliation{Center for Quantum Spintronics, Department of Physics, Norwegian University of Science and Technology, NO-7491 Trondheim, Norway}


\date{\today}

\begin{abstract}
	We consider an antiferromagnetic insulator that is in contact with a metal. Spin accumulation in the metal can induce spin-transfer torques on the staggered field and on the magnetization in the antiferromagnet. These torques relate to spin pumping: the emission of spin currents into the metal by a precessing antiferromagnet. We investigate how the various components of the spin-transfer torque are affected by spin-independent disorder and spin-flip scattering in the metal. Spin-conserving disorder reduces the coupling between the spins in the antiferromagnet and the itinerant spins in the metal in a manner similar to Ohm's law. Spin-flip scattering leads to spin-memory loss with a reduced spin-transfer torque. We discuss the concept of a staggered spin current and argue that it is not a conserved quantity. Away from the interface, the staggered spin current varies around a zero mean in an irregular manner. A network model explains the rapid decay of the staggered spin current.
\end{abstract}

\maketitle


\section{Introduction}

Charge currents cannot flow through antiferromagnetic insulators (AFIs). However, recent experiments have demonstrated that typical AFIs such as NiO and CoO are good spin conductors.\cite{Hahn:EPL2014,Wang:prl2014,Wang:prb2015,Moriyama:apl2015,LinPRL2016} One of the origins of these features is that spin pumping and spin transfer are as efficient across AFI-normal metal (NM) interfaces as in ferromagnet-NM systems.\cite{Cheng:prl2014} This potent spin transfer can empower low-dissipation high-frequency spin circuits in AFIs. \cite{Cheng:prl2016} In insulators, lossy itinerant charge carriers do not contribute to dissipation. 

These developments in insulators increase the potential applicability of antiferromagnetic spintronics. Antiferromagnets have no stray fields,\cite{Marti2014NatMat,marti2015prospect} and this feature might enable denser antiferromagnetic elements in future devices. However, the notable advantage of antiferromagnets compared to ferromagnets is that they can operate at speeds that are a hundred times faster \cite{Gomonay:prb2010,Cheng:prl2016,Gomonay:prl2016}. By using antiferromagnets, we can envision circuits that function in the largely unexplored spintronics THz regime. For instance, one can envision THz spin-torque oscillators.\cite{Cheng:prl2016,gomonay2012spin,khymyn2017antiferromagnetic,Slavin2017Arxiv} Furthermore, achieving such rapid spin dynamics is possible even in the absence of external magnetic fields.

Many of the phenomena that occur in ferromagnets also occur in antiferromagnets.\cite{Jungwirth:natnano2016,gomonay2014spintronics,baltzREVIEW2017} In antiferromagnetic metals and junctions involving antiferromagnets, there is a significant anisotropic magnetoresistance.\cite{Marti2014NatMat,Marti:prl2012,PhysRevB.95.064402_Jia_tunnel_MR_in_AF_junction,Park2011NatMat,SmejkalPRL2017} This magnetoresistance enables the detection of the staggered field.\cite{Marti:prl2012} There are also strong spin-orbit torques that can function to reorient the staggered field,\cite{Zelezny:prl2014,SmejkalPRL2017,PhysRevB.95.014403zelezny2017} as demonstrated in recent seminal experiments.\cite{Wadley:science2016,GrzybowskiPRL2017} 

In AFIs, spin waves can also reorient the staggered field.\cite{Tveten:prl2014,Gomonay:prl2016,benderPRL2017} In addition to spin-wave transport, spin superfluidity is also possible.\cite{Sonin:adp2010,Takei:prb2014} Spin superfluidity is analogous to superfluidity in helium-4 and could offer a low-dissipation route to spin transport.\cite{bunkovPRL2012,Takei:prl2014,Skarsvag:prl2015,AlirezaPRL2017_superfluid_NiO} 

In many of the phenomena that involve electrical control of spin transport in AFIs, spin transfer and spin pumping across AFI-metal interfaces are essential.\cite{Tserkovnyak2017PhysRevB.96.100402,Johansen2017PRBrapid,akash_kamra-ferri2017Arxiv,prb2017eirik:AFI-bose} Seminal papers have computed this interface spin coupling in the ideal case of no disorder and no spin-flip scattering within the metal\cite{Cheng:prl2014,Takei:prb2014}. The purpose of the present paper is to elucidate in further detail how the torques on the staggered field and on the magnetization occur and how these torques are influenced by various types of disorder.

To this end, we first reformulate the previous theories in a quantum language. We compute the rate of change of the spins at the $A$ and $B$ sublattices. In three-dimensional systems, where the number of involved spins is large, we evaluate the quantum mechanical rates of change in the classical limits of large spins and recover the previous results. Subsequently, we consider how the coupling between the spins in the AFI and the itinerant spins in the NM are affected by spin-conserving disorder and spin-flip-inducing disorder. 

The remainder of our manuscript is organized as follows. In section (\ref{sec:theory}), we present our model for the electrons in the NM, the spins in the AFI, and the coupling between the two sub-systems. In section (\ref{sec:numerical_results}), we present our numerical results of the spin-transfer torques (STTs) on the AFI and the spin currents in the NM. Finally, we discuss and summarize our findings in section (\ref{sec:conclusions}). 


\section{Spin transfer and spin pumping}
\label{sec:theory}
Dynamic antiferromagnets can pump spins into adjacent conductors even when they are insulating. Antiferromagnets are also affected by STTs arising from spin accumulations in neighboring metals. We will investigate how spin-conserving impurity scattering and spin-flip scattering influence these processes. Using a dynamic gauge transformation, we will relate spin pumping to spin transfer, similar to the Larmor theorem for ferromagnets.

\subsection{Spin-transfer torques}

\begin{figure}[h]
	\centering
	\includegraphics[width=0.7\linewidth]{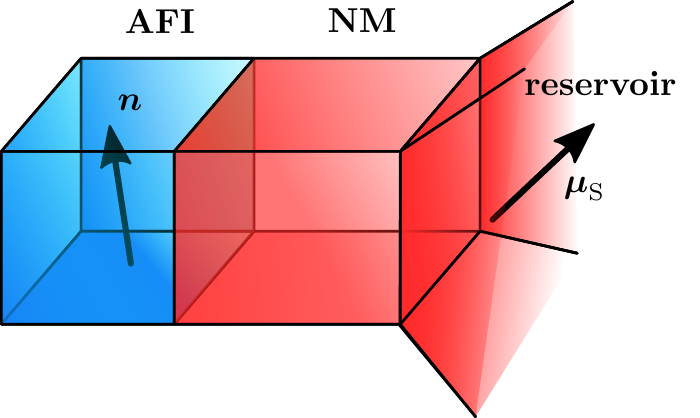}
	\caption{An antiferromagnetic insulator (AFI) is in contact with a normal metal (NM). The NM is connected to a reservoir with a spin accumulation, $\bs{\mu}_{\mr{S}}$.}
	\label{Fig:schematic}
\end{figure}

The AFI is located next to a NM (see Fig.\ \ref{Fig:schematic}).  A tight-binding model describes the itinerant electrons. At the interface, we assume that the AFI spins are exchange coupled to the itinerant spins in the metal. We assume that the spin dynamics in the AFI is spatially homogeneous. The spins behave as macrospins, albeit on two different sublattices. Excluding the spins in the bulk of the AFI, the Hamiltonian that involves the electron degrees of freedom is 
\begin{equation}
\hat{H} = \hat{H}_{\mr{NM}} + \hat{H}_{\mathrm{AF,N}} \, , 
\end{equation}
where $\hat{H}_{\mr{NM}}$ describes the NM and $\hat{H}_{\mathrm{AF,N}}$ characterizes the coupling between the itinerant spins and the localized AFI spins. 

The Hamiltonian for the itinerant electrons in the NM reads as
\begin{equation}
\hat{H}_{\mr{NM}}=-\tilde{t}\sum_{\langle\bs{r},\bs{r}'\rangle}\hat{c}^{\dagger}_{\bs{r}}\hat{c}_{\bs{r}'}+6\tilde{t}\sum_{\bs{r}}\hat{c}^{\dagger}_{\bs{r}}\hat{c}_{\bs{r}}  +\sum_{\bs{r}}\hat{c}^{\dagger}_{\bs{r}}V_{\bs{r}}\hat{c}_{\bs{r}} \, ,
\label{eq:HtotNM}
\end{equation}
where $\hat{c}^{\dagger}_{\bs{r}}=\big(\begin{array}{cc}\hat{c}^{\dagger}_{\bs{r}\uparrow} \, ,  &\hat{c}^{\dagger}_{\bs{r}\downarrow}\end{array}\big)$ is in terms of the creation (annihilation) operator $\hat{c}^{\dagger}_{\bs{r}s}$ $ (\hat{c}_{\bs{r}s})$ of an electron at site $\bs{r}$ with spin projection $s=\uparrow$ or $s=\downarrow$ along the spin-quantization axis $\hat{z}$. In Eq.\ \eqref{eq:HtotNM}, the sum ($\left\langle\bs{r},\bs{r}' \right\rangle$) is over nearest neighbors and $\tilde{t}$ is the hopping energy.
The onsite potential consists of a spatially constant contribution, $6\tilde{t}$, and a random potential, $V_{\bs{r}}$, that models disorder. We will specify the statistical properties of the random potential below. We set the lattice constant as $a=1$. Operators and unit vectors are denoted by a hat ($~\hat{}~$).

Even in antiferromagnets where the spins precess rapidly at THz frequencies, the spin dynamics are slower than the electron dynamics. The STT on the AFI can therefore be computed as a scattering problem that keeps the localized spins {\it static}. We assume that the NM couples to a large reservoir where there may be spin and charge accumulations. The out-of-equilibrium distribution of the incoming electrons is therefore assumed to be known in the reservoirs. The out-of-equilibrium spin and charge densities are governed by the time-independent scattering matrix $S$. In our geometry, there is only one lead. The scattering matrix thus only includes reflection matrices.

\begin{figure}[h]
	\centering
	\includegraphics[width=0.9\columnwidth]{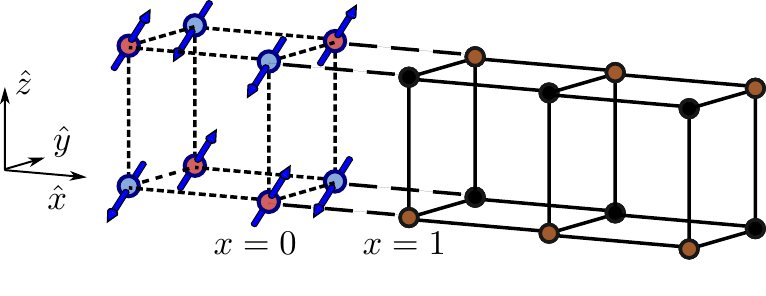}
	\caption{Sections of the AFI and the NM around their interface. In the AFI ($x\leq0$), the localized spins residing on the red (blue) nodes are at sublattice $\mc{A}$ ($\mc{B}$). The localized spins at the interface ($x=0$) are exchange coupled to adjacent itinerant electron spins ($x=1$), as indicated by semi-long dashes. In the NM ($x\geq1$), we use a different convention for the sublattices. In the NM, the brown (black) sites reside on sublattice $A$ ($B$). The solid black lines illustrate the itinerant electron hopping.  }
	\label{Fig:AFI-NM-exchange-coupling}
\end{figure}

Our focus is on the spin information transferred between the AFI and the NM. In the NM, the itinerant spin density operator at position $\bs{r}$ is 
\begin{equation}
\bs{\mathfrak{\hat{s}}}_{\bs{r}}=\frac{\hbar}{2}\hat{c}^{\dagger}_{\bs{r}}\bs{\sigma}\hat{c}_{\bs{r}} \, ,
\end{equation}
in terms of the $2\times 2$ spin-space vector of Pauli matrices, $\bs{\sigma}=(\sigma_x,\sigma_y,\sigma_z)$.
We separate the longitudinal, $x$, and transverse, $\bs{r}_{\mperp}=(y,z)$,  coordinates, $\bs{r}\rightarrow (x,\bs{r}_{\mperp})$. Furthermore, we use a discrete index to label the lattice position along the $\hat{x}$-direction such that $x \rightarrow 0,1,\ldots$. For instance, $\hat{\bs{\mf{s}}}_{1,\bs{r}_{\mperp}}$ denotes the electron spins in the transverse layer $x=a$. The interface exchange coupling between the itinerant electron spins at the interface to the adjacent localized AFI spins is (see Fig.\ \ref{Fig:AFI-NM-exchange-coupling})
\begin{equation}
\hat{H}_{\mr{AF,N}}=J\sum_{\bs{r}_{\mperp}}\bs{\mc{\hat{S}}}_{0,\bs{r}_{\mperp}}\cdot\bs{\mf{\hat{s}}}_{1,\bs{r}_{\mperp}} \, ,
\end{equation}
where $J$ is the interface exchange coupling. $\bs{\mc{\hat{S}}}_{0,\bs{r}_{\mperp}}$ is the operator associated with the localized spin in the AFI next to the NM. 

The spin current in the longitudinal ($\hat{x}$) direction is
\begin{align}
\bs{\hat{j}}^{\pmm\hat{x}}_{\bs{r}}=\frac{i\tilde{t}}{2}\left(\hat{c}^{\dagger}_{\bs{r}\pm\bs{\delta}_{x}}\bs{\sigma}\hat{c}_{\bs{r}}-\hat{c}^{\dagger}_{\bs{r}}\bs{\sigma}\hat{c}_{\bs{r}\pm\bs{\delta}_{x}} \right) \, .
\label{eq:spincurrentx}
\end{align}
In Eq.\ \eqref{eq:spincurrentx}, the superscript $+(-)$ classifies spin currents to the right (left) when electrons hop from $x$ to $x+a$ ($x$ to $x-a$). $\bs{\delta}_{x}=a\hat{x}$ equals one lattice constant in the $\hat{x}$-direction. Similarly, we define the spin currents in the two transverse directions: $\bs{\hat{j}}^{\pmm\hat{y}}_{\bs{r}}$ and $\bs{\hat{j}}^{\pmm\hat{z}}_{\bs{r}}$. Correspondingly, the hoppings are $\bs{\delta}_{y}$ and $\bs{\delta}_{z}$ in the $\hat{y}$- and $\hat{z}$-directions, respectively. 

We consider a bipartite AFI. A localized spin on sublattice $\mc{A}$ has six nearest-neighbor spins on sublattice $\mc{B}$ and vice versa. The lattice is cubic with lattice constant $2a$ for sublattice $\mc{A}$. Sublattice $\mc{B}$ is displaced by $a\hat{x}$ with respect to sublattice $\mc{A}$. 

For convenience, we use a {\it different} set of sublattices in the NM, but we also label these sublattices as sublattices $A$ and $B$. This notation describes the (staggered) spin currents in the most transparent way. In our coordinate system, sublattice $A$ is spanned by the vectors $\{\bs{r}_{A}\} = \{ (a,0,0), (0, a ,a), (0,a,-a)\} $. Sublattice $B$ is displaced by $a\hat{y}$ with respect to sublattice $A$, as shown in Fig.\ \ref{Fig:AFI-NM-exchange-coupling}. The volume of a unit cell in sublattice $A$ or $B$ is then $2a^{3}$.

We separate the currents into terms that are associated with the rate of change of the spin at the $A$ and $B$ sublattices. The spin current in the longitudinal direction associated with lattice sites $A$ in layer $x$ is  
\begin{equation}
\bs{\mathcal{\hat{I}}}^{\mr{s},A}_{x}=\begin{cases}
\sum_{\bs{r}_{\mperp}\in A}\bs{\hat{j}}^{\smm\hat{x}}_{x,\bs{r}_{\mperp}} &  x=a \, , \\
\sum_{\bs{r}_{\mperp}\in A}\frac{1}{2}\left(\bs{\hat{j}}^{\smm\hat{x}}_{x,\bs{r}_{\mperp}}-\bs{\hat{j}}^{\sm\hat{x}}_{x,\bs{r}_{\mperp}}\right) & x>a \, .
\end{cases}
\end{equation}
We define a similar relation for the spin current in the longitudinal direction associated with lattice sites $B$,  $\bs{\mathcal{\hat{I}}}^{\mr{s},B}_{x}$. The total longitudinal spin current at layer $x$ is the total spin current carried at both sublattices, $\bs{\mc{\hat{I}}}^{\mr{s}}_{x}=\bs{\mc{\hat{I}}}^{\mr{s},A}_{x}+\bs{\mc{\hat{I}}}^{\mr{s},B}_{x}$. 

\begin{figure}[h]
	\centering
	\includegraphics[width=0.8\columnwidth]{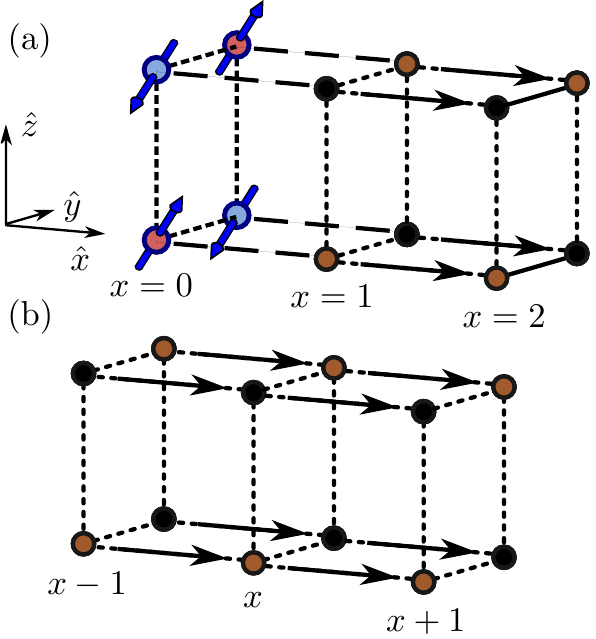}
	\caption{(a) The spin current $\bs{\mc{I}}^{\mr{s}}_{1}$ at the interface ($x=1$), indicated by black arrows, determines the torque on the magnetization $\bs{\tau}_{\mc{M}}$. (b) The spin current $\bs{\mc{I}}^{\mr{s}}_{x}$ in the NM bulk ($x>1$) is the average of the spin current between the layers at $x-1$ and $x$ and between the layers at $x$ and $x+1$.  }
	\label{Fig:SpinCurrent}
\end{figure}

The NM has finite lengths $N_{y}a$ and $N_{z}a$ in the transverse directions $\hat{y}$ and $\hat{z}$, where $N_{y}$ and $N_{z}$ are the number of lattice sites, respectively. In the transverse direction $\hat{y}$, the spin currents (between the sublattices) are 
\begin{equation}
\bs{\mathcal{\hat{I}}}^{\Delta \hat{y}}_{x,\bs{r}_{\mperp}}=\begin{cases}
\bs{\hat{j}}^{\smm\hat{y}}_{\bs{r}} & y=1 \, , \\
\bs{\hat{j}}^{\smm\hat{y}}_{\bs{r}}-\bs{\hat{j}}^{\sm\hat{y}}_{\bs{r}} & 1<y<N_{y} \, , \\
-\bs{\hat{j}}^{\sm\hat{y}}_{\bs{r}} & y=N_{y} \, .
\end{cases}
\end{equation}
A similar expression is obtained for the spin current in the transverse direction $\hat{z}$, $\bs{\mc{\hat{I}}}^{\Delta \hat{z}}_{x,\bs{r}_{\mperp}}$, by substituting $y\rightarrow z$.

\begin{figure}[h]
	\centering
	\includegraphics[width=0.8\columnwidth]{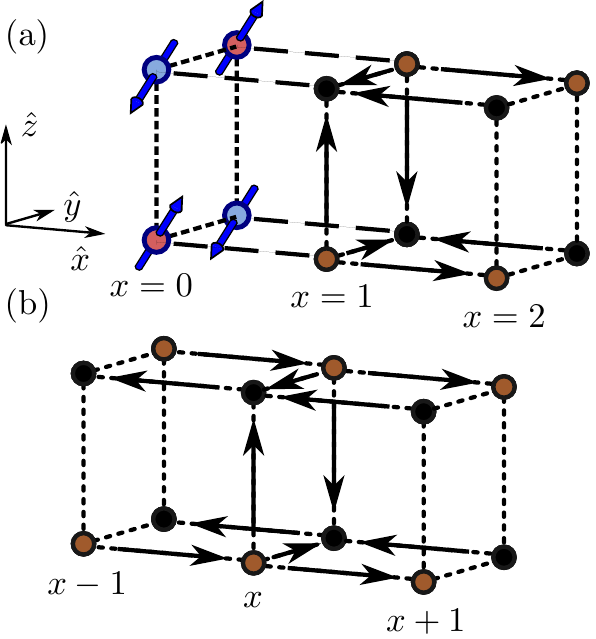}
	\caption{(a) The staggered spin current $\bs{\mathcal{I}}^{\mathrm{ss}}_{1}$ at the AFI-NM interface ($x=1$), which is shown by black arrows. The torque $\bs{\tau}_{\mc{N}}$ on the Neel order parameter is determined by $\bs{\mathcal{I}}^{\mathrm{ss}}_{1}$. (b) The staggered spin current $\bs{\mathcal{I}}^{\mathrm{ss}}_{x}$ at layer $x>1$ includes spin currents between lattices $A$ and $B$ within the layer at $x$ (the interlayer staggered transverse current $\bs{\mc{I}}^{AB}_{x}$) and longitudinal staggered spin currents between the layers at $x-1$ and $x$ and between the layers at $x$ and $x+1$. Note the factor of 2 in front of $\bs{\mc{I}}^{AB}_{x}$ in the definition of the staggered spin current  $\bs{\mc{I}}^{\mr{ss}}_{x}$. }
	\label{Fig:CubicLatticeStaggeredCurrent}
\end{figure}

A staggered spin current involves the transfer of spin angular momentum between the two sublattices within each transverse layer. We define the interlayer staggered transverse current operator as
\begin{equation}
\bs{\mathcal{\hat{I}}}^{AB}_{x}=\sum_{\bs{r}_{\mperp}\in A}\left(\bs{\mathcal{\hat{I}}}^{\Delta \hat{y}}_{x,\bs{r}_{\mperp}}+\bs{\mathcal{\hat{I}}}^{\Delta z}_{x,\bs{r}_{\mperp}}\right)-\sum_{\bs{r}_{\mperp}\in B}\left(\bs{\mathcal{\hat{I}}}^{\Delta \hat{y}}_{x,\bs{r}_{\mperp}}+\bs{\mathcal{\hat{I}}}^{\Delta z}_{x,\bs{r}_{\mperp}}\right) \, ,
\end{equation}
as illustrated in Fig.\ \ref{Fig:CubicLatticeStaggeredCurrent}. The staggered spin current at layer $x$ is then
\begin{align}
\bs{\mathcal{\hat{I}}}^{\mathrm{ss}}_{x}=\bs{\mathcal{\hat{I}}}^{\mathrm{s},A}_{x}-\bs{\mathcal{\hat{I}}}^{\mathrm{s},B}_{x}+2\bs{\mathcal{\hat{I}}}^{AB}_{x} \, ,
\label{eq:staggSpinCurrentInLayerX}
\end{align}
which consists of a longitudinal part and a transverse part.

We are interested in the average torque per spin in the AFI. The total number of localized spins is $N_{\mr{AF}}=N_{\perp}N_{\parallel}$, where $N_{\perp}$ is the number of spins in one transverse layer and $N_{\parallel}$ is the number of transverse layers in the longitudinal direction. The total volume of the AFI is then $N_{\mr{AF}}a^{3}$. We define the total spin of the AFI as
\begin{equation}
\bs{\mc{S}}_{\mc{M}}=\sum_{\bs{r}}\langle\bs{\mc{\hat{S}}}_{\bs{r}}\rangle \, ,
\label{eq:SM}
\end{equation}
where the brackets denote the expectation value. 

The difference between the total spin on sublattices $A$ and $B$ is
\begin{equation}
\bs{\mc{S}}_{\mc{N}}=\sum_{\bs{r}\in A}\langle\bs{\mc{\hat{S}}}_{\bs{r}}\rangle-\sum_{\bs{r}\in B}\langle\bs{\mc{\hat{S}}}_{\bs{r}}\rangle \, .
\label{eq:SN}
\end{equation}
The total spin $\bs{\mc{S}}_{\mc{M}}$ and the spin difference $\bs{\mc{S}}_{\mc{N}}$ are related to the total magnetization and the Neel order parameter in the AFI, respectively. We define the corresponding torques per spin $\bs{\tau}_{\mc{M}}=\partial_{t}\bs{\mc{S}}_{\mc{M}}/N_{\mr{AF}}$ and $\bs{\tau}_{\mc{N}}=\partial_{t}\bs{\mc{S}}_{\mc{N}}/N_{\mr{AF}}$.

In the steady state, the itinerant spins do not vary in time, $\left\langle \partial_{t}\hat{\bs{\mf{s}}}_{\bs{r}} \right\rangle=0$. The rate of change of the localized spins can thus be expressed in terms of steady-state spin currents in the NM via the spin continuity equations (Eqs.\ \eqref{eq:localizedSpinsRateOfChange} and \eqref{eq:itinerantspincontinuity}). The torques on the AFI are
\begin{subequations}\label{subeqs:torquesAsSpinCurrents}
	\begin{align}
	\bs{\tau}_{\mc{M}}=&-\frac{1}{N_{\mathrm{AF}}}\bs{\mc{I}}^{\mr{s}}_{1} \, , \label{eq:torqueMagnetizationAsSpinCurrent} \\
	\bs{\tau}_{\mc{N}}=&-\frac{1}{N_{\mathrm{AF}}}\bs{\mathcal{I}}^{\mathrm{ss}}_{1} \, , \label{eq:torqueNeelAsStaggSpinCurrent}
	\end{align}
\end{subequations}
in terms of the expectation values of the spin current and the staggered spin current at the interface (layer $x=1$).

\subsection{Umklapp scattering}\label{subsubsection_umklapp}

In the lead, scattering states coupled to out-of-equilibrium reservoirs  determine the torques on the AFI. Energy is conserved and the transport is elastic. The scattering states are labeled by the incoming (transverse) mode $n$, energy $E$, and spin $\alpha$:
\begin{align}
\psi_{nE\alpha}(\bs{r}) =& \frac{1}{\sqrt{h v_{n}}}\xi_{\alpha}\varphi_{n}(\bs{r}_{\mperp})e^{-ik^{\parallel}_{n}x}\nonumber \\
&+\sum_{m=1}^{N_{\mperp}}\sum_{\beta=\uparrow,\downarrow}\frac{1}{\sqrt{h v_{m}}}r^{\beta\alpha}_{mn}\xi_{\beta}\varphi_{m}(\bs{r}_{\mperp})e^{ik^{\parallel}_{m}x} \, ,
\label{eq:DefinitionOfScatteringState}
\end{align}
where $\hbar v_{n}=\partial E/\partial k^{\parallel}_{n}$ is the group velocity of the propagating wave and $h=2\pi\hbar$ is Planck's constant. $\xi_{\alpha}$ is a spinor,  $\xi^{\dagger}_{\uparrow} = (\begin{array}{cc}1   &0\end{array})$, and $\xi^{\dagger}_{\downarrow} = (\begin{array}{cc}0   &1\end{array})$. The transverse wavefunction $\varphi_{n}(\bs{r}_{\mperp})$ is an eigenfunction of the transverse part of $\hat{H}_{\mr{NM}}$ with eigenenergy $\varepsilon^{\mperp}_{n}\in[-4\tilde{t},4\tilde{t}]$.  The longitudinal wavenumber is positive, $k^{\parallel}_{n}>0$; thus, $e^{\mp ik^{\parallel}_{n}x}$ propagates to the left (-) or to the right (+). The scattering state has a total energy of
\begin{align}
E = \varepsilon^{\mperp}_{n} + 2\tilde{t} \left(3-\cos k^{\parallel}_{n}a\right) \, .
\end{align}
The total number of transverse modes is $N_{\mperp}$. A mode is propagating if its energy $E$ lies within its band, i.e., if  $\varepsilon^{\mperp}_{n}+4\tilde{t}< E <\varepsilon^{\mperp}_{n}+8\tilde{t}$. The number of propagating modes is then $N_{\mr{p}}<N_{\mperp}$. We will consider a NM at half-filling such that the Fermi energy is $E_{\mr{F}} = 6\tilde{t}$.  The reflection amplitudes $r^{\beta\alpha}_{mn}$ describe reflection from incoming mode $n$ to outgoing mode $m$ and from spin $\alpha$ to $\beta$. 

When there is no spin-flip scattering, the reflection matrix in Eq.\ \eqref{eq:DefinitionOfScatteringState} simplifies to
\begin{equation}
r^{\beta\alpha}_{mn}= \delta_{\alpha\beta}\frac{1}{2}(r^{\uparrow}_{mn}+r^{\downarrow}_{mn})+\frac{1}{2}(r^{\uparrow}_{mn}-r^{\downarrow}_{mn})\bs{n}\cdot\bs{\sigma}_{\beta\alpha} \, ,
\end{equation}
in terms of reflection coefficients for spin up (down), $r^{\uparrow}_{mn}$ ($r^{\downarrow}_{mn}$), along the direction of the Neel order parameter.  

The scattering states constitute a current-normalized and complete set of states. Therefore, the reflection matrix is unitary, and particle current is conserved. The electron field operator (inside the lead) is expressed in terms of the scattering states as
\begin{equation}
\hat{c}_{\bs{r}}(t) = \sum_{n=1}^{N_{\mperp}}\sum_{\alpha=\uparrow,\downarrow}\int_{\varepsilon^{\mperp}_{n}}^{\infty} dE\psi_{nE\alpha}(\bs{r})\hat{c}_{n\alpha}(E)e^{-iEt/\hbar} \, ,
\label{eq:DefinitionOfTheFieldOperator}
\end{equation}
where we define the operator $\hat{c}_{n\alpha}(E)$ ($\hat{c}^{\dagger}_{n\alpha}(E)$), which annihilates (creates) the scattering state $\psi_{nE\alpha}$ originating from the reservoir.\cite{buttiker1992scatteringPRB} 

The distribution function in the reservoir determines the expectation values of the operators. We allow for a small spin accumulation in the reservoir. Then, 
\begin{equation}
\left\langle\hat{c}^{\dagger}_{n\alpha}(E)\hat{c}_{m\beta}(E')  \right\rangle =\delta_{nm}\delta(E-E')f_{\beta\alpha}(E) \, .
\label{eq:expectationValueReservoir}
\end{equation}
The distribution function is 
\begin{equation}
f_{\alpha\beta}(E) = \delta_{\alpha\beta}f_{\mr{FD}}(E-\mu) + \bs{\sigma}_{\alpha\beta}\cdot \bs{f}_{\mr{S}}(E) \, ,
\label{eq:distributionFunction}
\end{equation}
in terms of the Fermi-Dirac function $f_{\mr{FD}}(E-\mu)$ with chemical potential $\mu$ and temperature $T$. The function $\bs{f}_{\mr{S}}(E)$ describes the spin accumulation $\bs{\mu}_{\mr{S}} = \int dE\mr{Tr}_{\mr{s}}\{\bs{\sigma}(\bs{\sigma}\cdot\bs{f}_{\mr{S}}(E))\}$ by taking the trace in spin space.\cite{arne2000PhysRevLett.84.2481}

It is illustrative to first consider the ideal case of no disorder and no spin-flip scattering, as considered previously \cite{Cheng:prl2014,Takei:prb2014}. For simplicity, we also assume periodic boundary conditions in the transverse directions. There are $N$ lattice sites in both transverse directions ($\hat{y}$ and $\hat{z}$). The orthonormal transverse wavefunctions are $\varphi_{n}(\bs{r}_{\mperp})\rightarrow \varphi_{n_{y}}(y)\varphi_{n_{z}}(z) $, where
\begin{equation}
\varphi_{n_{y}}(y) = \frac{1}{\sqrt{N}}e^{2\pi iyn_{y}/Na} \, ,
\end{equation}
with discrete values $n_{y}= 1,2, ..., N$ and similarly for the wave function $\varphi_{n_{z}}(z)$ that is characterized by $n_{z}$. The resulting scattering matrix consists of normal scattering
\begin{equation}
\frac{1}{2}(r^{\uparrow}_{m_{y}m_{z},n_{y}n_{z}}+r^{\downarrow}_{m_{y}m_{z},n_{y}n_{z}}) = A_{n_{y}n_{z}}\delta_{m_{y}n_{y}}\delta_{m_{z}n_{z}} \, ,
\end{equation}
and spin-dependent umklapp scattering
\begin{equation}
\frac{1}{2}(r^{\uparrow}_{m_{y}m_{z},n_{y}n_{z}}-r^{\downarrow}_{m_{y}m_{z},n_{y}n_{z}}) = B_{n_{y}n_{z}}\delta_{m_{y}\bar{n}_{y}}\delta_{m_{z}\bar{n}_{z}} \, ,
\label{scatteringMatrixTermUmklapp}
\end{equation}
with coefficients $A_{n_{y}n_{z}}$ and $B_{n_{y}n_{z}}$ defined in the Appendix in Eqs.\ \eqref{eq:normalScattCoeff} and \eqref{eq:umklappScattCoeff}. Umklapp scattering changes the incoming transverse modes from $(n_{y},n_{z})$ to $(\bar{n}_{y},\bar{n}_{z})$ with four possible combinations: $(\bar{n}_{y},\bar{n}_{z}) = (n_{y}\pm\frac{N}{2},n_{z}\pm\frac{N}{2})$ or $(\bar{n}_{y},\bar{n}_{z}) =(n_{y}\pm\frac{N}{2},n_{z}\mp\frac{N}{2})$. When  the incoming mode $n_{y}$ is in the range from $1$ to $\frac{N}{2}$, then $\bar{n}_{y} = n_{y} +\frac{N}{2}$, and when $n_{y}$ is in the range from $\frac{N}{2}+1$ to $N$, then $\bar{n}_{y} = n_{y} -\frac{N}{2}$. There are similar considerations for $n_{z}$.  

A finite magnetization may induce spin precession of the conduction electrons. To separate the different contributions to the torques in a transparent manner, we focus on AFIs with a vanishingly small magnetization. In clean systems, we compute that the torques on an AFI with volume $\mc{V}_{\mr{AF}} = N_{\mr{AF}}a^{3}$ are 
\begin{subequations}\label{eq:the_torques}
	\begin{align}
	\bs{\tau}_{\mc{M}}=& \frac{1}{N_{\mr{AF}}} \frac{1}{4\pi}g^{\mperp}_{\mathrm{m}} \bs{n}\times(\bs{\mu}_{\mr{S}}\times\bs{n}) \, ,\\
	\bs{\tau}_{\mc{N}}=&\frac{1}{N_{\mr{AF}}}\frac{1}{4\pi}g^{\mperp}_{\mr{n}}(\bs{\mu}_{\mr{S}}\times\bs{n}) \, ,
	\end{align}
\end{subequations}
in terms of the dimensionless spin-mixing conductances 
\begin{subequations}\label{eq:spinmix_conductances}
	\begin{align}
	g^{\mperp}_{\mr{m}} =& \frac{8\lambda^{2}N_{\perp}}{(1+\lambda^{2})^{2}} \sum_{n_{y}n_{z}}  \sin^{2}k^{\parallel}_{n_{y}n_{z}}a    \, ,\label{eq:spinmix_magnetiazation} \\
	g^{\mperp}_{\mr{n}} =& \frac{4\lambda N_{\perp} }{ (1+\lambda^{2})^{2}}\sum_{n_{y}n_{z}}(-\sin k^{\parallel}_{n_{y}n_{z}}a)\cos 2k^{\parallel}_{n_{y}n_{z}}  a  \, , \label{eq:spinmix_neel} 
	\end{align}
\end{subequations}
where the longitudinal wavenumber $k^{\parallel}_{n_{y}n_{z}}$ is evaluated at the Fermi energy $E_{\mr{F}}$. The summations are over the fraction of the Brillouin zone (BZ) where the modes are propagating. $\lambda = J\mc{S}/\tilde{t}$ parametrizes the exchange coupling at the interface, where $\mc{S}$ is the localized spin. The torque $|\bs{\tau}_{\mc{M}}|$ obtains its maximum for $\lambda =\pm 1$, whereas $|\bs{\tau}_{\mc{N}}|$ has the highest value at $\lambda = \pm\frac{1}{\sqrt{3}}$. Without the $\lambda$-dependent fractions in Eqs.\ \eqref{eq:spinmix_magnetiazation} and \eqref{eq:spinmix_neel}, the BZ summations of the trigonometric functions in $g^{\mperp}_{\mr{m}}$ and $g^{\mperp}_{\mr{n}}$ yield values on the order of unity (numerical values of $\approx 4.57862$ and $\approx 3.163$, respectively).

Note that there is an important difference in the definitions of the spin-mixing conductances $g^{\mperp}_{\mr{m}}$ and $g^{\mperp}_{\mr{n}}$ for the AFI (Eq.\ \eqref{eq:spinmix_conductances}) compared to the usual expression for the spin-mixing conductance \cite{arne2000PhysRevLett.84.2481} $g^{\uparrow\downarrow} = N_{\mr{p}} - \sum_{mn}r^{\uparrow}_{mn}(r^{\downarrow}_{mn})^{*}$ for ferromagnets. For the AFI discussed above, we find in the clean limit that $g^{\uparrow\downarrow} = \sum_{n_{y}n_{z}}2|B_{n_{y}n_{z}}|^{2}$ in terms of the amplitude for umklapp scattering $B_{n_{y}n_{z}}$ from the scattering matrix (Eqs.\ \eqref{scatteringMatrixTermUmklapp} and \eqref{eq:umklappScattCoeff}). The spin-mixing conductance $g^{\mperp}_{\mr{m}}$ can be expressed in terms of $g^{\uparrow\downarrow}$ as the real part: $g^{\mperp}_{\mr{m}} = \mr{Re}\left\{g^{\uparrow\downarrow} \right\}$.
The spin-mixing conductance $g^{\mperp}_{\mr{n}}$ is {\it not} related to $g^{\uparrow\downarrow}$ in a simple way. Instead, $g^{\mperp}_{\mr{n}}$ has to be calculated directly from the wavefunctions at the interface. Consequently, we find that $g^{\mperp}_{\mr{n}}$ arises from an expression that is a combination of normal and umklapp scattering coefficients, $A_{n_{y}n_{z}}$ and $B_{n_{y}n_{z}}$, which is then reduced to Eq.\ \eqref{eq:spinmix_neel}. 

\subsection{Spin pumping from an antiferromagnetic insulator}\label{section:theory}

We now consider the dynamic situation where the localized spins in the AFI rotate uniformly. Similar to how precessing spins in an FM pump a spin current into an adjacent metal, we show that a dynamical AFI pumps both a spin current and a staggered spin current. 

To demonstrate this property, we perform a time-dependent gauge transformation.  The spin axes rotate and transform the time-dependent Hamiltonian $H(t)$ in the lab frame to a static Hamiltonian $\tilde{H}$ with an effective magnetic field in the rotating frame. The same gauge transformation relates the spin currents in the lab frame to the spin currents in the rotating frame. Static scattering states determine the latter spin currents. 

In the classical limit of many AF spins, the coupling between the AFI and the metal is
\begin{equation}
\hat{H}_{\mr{AF,N}}(t)=J\frac{\hbar}{2}\sum_{\bs{r}_{\mperp}}\bs{\mc{S}}_{0,\bs{r}_{\mperp}}(t)\cdot\hat{c}^{\dagger}_{1,\bs{r}_{\mperp}}\bs{\sigma}\hat{c}_{1,\bs{r}_{\mperp}} \, ,
\label{eq:hamiltonian_dynamic_exchange}
\end{equation}
where the dynamic spins $\bs{\mc{S}}_{0,\bs{r}_{\mperp}}(t) = \pm\mc{S}\bs{n}(t)$ are parallel or anti-parallel to the direction of the Neel order parameter $\bs{n}(t)$. 

The gauge transformation rotates the spin space. In the rotating frame, the field operator is 
\begin{equation}
\hat{\tilde{c}}_{\bs{r}} = U(t)\hat{c}_{\bs{r}} \, , 
\end{equation}
where $U(t)$ is a time-dependent, unitary $2\times 2$ matrix, $U^{\dagger}=U^{-1}$. The evolution of $\hat{\tilde{c}}_{\bs{r}}(t)$ is governed by the Hamiltonian 
\begin{equation}
\tilde{H} = UHU^{\dagger} -i\hbar U\frac{\partial U^{\dagger}}{\partial t} \, , 
\end{equation}
which adds the gauge potential $V_{\mr{eff}} = -i\hbar U \partial_t U^{\dagger}$, as shown in App.\ \ref{appendix:gauge_transformation}. The time-dependent problem with the Hamiltonian $H(t)$ is transformed into a static problem by finding a unitary matrix $U(t)$ that makes the terms $U(\bs{n}(t)\cdot\bs{\sigma})U^{\dagger}$ and $V_{\mr{eff}}$ both independent of time.

\begin{figure}[h]
	\centering
	\includegraphics[width=0.75\columnwidth]{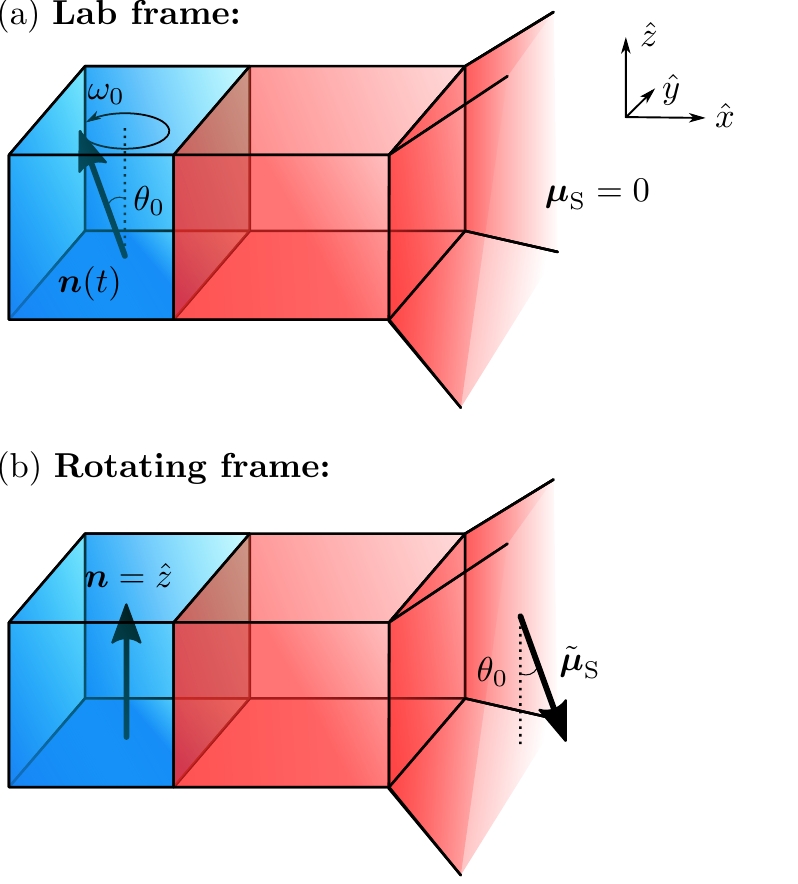}
	\caption{(a) Laboratory frame of reference, where the AFI spins precess around the $\hat{z}$-axis with frequency $\omega_{0}$. The Neel vector $\bs{n}(t)$ is at an angle $\theta_{0}$ with respect to the $\hat{z}$-axis. There is no spin accumulation in the lab frame. (b) In the rotating spin frame, the Neel vector points in the $\hat{z}$-direction. The gauge potential $V_{\mr{eff}}$ induces a finite spin accumulation $\bs{\tilde{\mu}}_{\mr{S}}$ in the NM, including the reservoir. }
	\label{Fig:labframe_vs_rotatingframe}
\end{figure}

A time-dependent gauge transformation can describe spin pumping, both in a ferromagnet \cite{tserkovnyak2005RevModPhys.77.1375,Tatara2016PhysRevB.94.224412} and in an antiferromagnet, when the spins precess around a static precession axis or if the precession axis changes slowly compared to the precession frequency. We now illustrate this property by considering a Neel order parameter $\bs{n}(t) = \hat{x}\cos\omega_{0}t\sin\theta_{0}+\hat{y}\sin\omega_{0}t\sin\theta_{0}+\hat{z}\cos\theta_{0}$ that rotates around the $\hat{z}$-axis. The spins precess with angular frequency $\omega_{0}$ and a constant polar angle $\theta_{0}$, as shown in Fig\ \ref{Fig:labframe_vs_rotatingframe}. 
The spin-rotation matrix that yields a static Hamiltonian in the rotating frame is then 
\begin{equation}
U(t)=e^{i\sigma_{y}\theta_{0}/2}e^{i\sigma_{z}\omega_{0} t/2} \, .
\end{equation}
$U(t)$ diagonalizes $U(\bs{n}(t)\cdot\bs{\sigma})U^{\dagger} = \sigma_{z}$. The resulting gauge potential is
\begin{equation}
V_{\mr{eff}} = \frac{\hbar\omega_{0}}{2}(\sin\theta_{0}\sigma_{x}-\cos\theta_{0}\sigma_{z}) \, .
\end{equation}
A Zeeman-like gauge potential in the NM reservoir splits the population of spins and is equivalent to a spin accumulation 
\begin{equation}
\bs{\tilde{\mu}}_{\mr{S}} = \hbar\omega_{0}(\sin\theta_{0}\hat{x}-\cos\theta_{0}\hat{z}) \, . 
\end{equation}
We now consider the spin currents in the rotating frame. Here, the static AFI spins point along $\hat{z}$. In linear response, we can calculate the spin currents by disregarding the gauge-transformation-induced magnetic field in the scattering region while retaining a spin accumulation in the form of $\bs{\tilde{\mu}}_{\mr{S}}$ in the reservoir. Only the component $\bs{\tilde{\mu}}_{\mr{S}}\cdot\hat{x}=\hbar\omega_{0}\sin\theta_{0}$ results in the non-zero spin current $\bs{\mc{\tilde{I}}}^{\mr{s}}_{1}=(1/4\pi)g^{\mperp}_{\mr{r}}\hbar\omega_{0}\sin\theta_{0}(-\hat{x})$ and the staggered spin current $\bs{\mathcal{\tilde{I}}}^{\mathrm{ss}}_{1}= (1/4\pi)g^{\mperp}_{\mr{i}}\hbar\omega_{0}\sin\theta_{0}\hat{y}$. Spin currents in the rotating frame (denoted with a tilde) are defined similar to Eq.\ \eqref{eq:spincurrentx} with $\hat{c}_{\bs{r}}\rightarrow\hat{\tilde{c}}_{\bs{r}}$. When transforming back to the lab frame, we use that 
\begin{equation}
\hbar \omega_{0}\sin\theta_{0}U^{\dagger}(-\sigma_{x})U = \hbar (\bs{n}\times\bs{\dot{n}})\cdot\bs{\sigma}\, ,
\end{equation}
and $\hbar\omega_{0}\sin\theta_{0}U^{\dagger}\sigma_{y}U = \hbar \bs{\dot{n}}\cdot\bs{\sigma}$, such that the spin currents in the lab frame are 
\begin{subequations}
	\begin{align}
	\bs{\mc{I}}^{\mr{s}}_{1} & =\frac{1}{4\pi}g^{\mperp}_{\mr{r}}\bs{n}\times\bs{\dot{n}} \, ,\label{eq:pumped_spin_current} \\ \bs{\mathcal{I}}^{\mathrm{ss}}_{1} & =\frac{1}{4\pi} g^{\mperp}_{\mr{i}}\bs{\dot{n}} \, . \label{eq:pumped_staggered_spin_current}
	\end{align}
\end{subequations}
The expressions for the pumped spin currents in Eqs. \eqref{eq:pumped_spin_current} and \eqref{eq:pumped_staggered_spin_current} are consistent with the results obtained from a time-dependent scattering-matrix approach \cite{Cheng:prl2014} in the limit of a vanishing magnetization, $\bs{m}\rightarrow 0$.

In other words, the spin current $\bs{\mc{I}}^{\mr{s}}_{1}$ and the staggered spin current $\bs{\mathcal{I}}^{\mathrm{ss}}_{1}$ pumped from a precessing Neel order parameter can be calculated by considering a static problem, where the Neel order parameter points along the $\hat{z}$-axis with a spin accumulation $\hbar\omega_{0}\sin\theta_{0}\sigma_{x}\hat{x}$ in the reservoir. 

\subsection{Decay of a staggered current}\label{section:decay_stagg_current}
The staggered spin current $\bs{\mc{I}}^{\mr{ss}}_{1}$ has a finite value at the AFI-NM interface, which induces a torque on the Neel order parameter. However, in the NM, the staggered spin current $\bs{\mc{I}}^{\mr{ss}}_{x}$ is expected to decay rapidly away from the interface because it is not conserved inside the NM. We now illustrate in a heuristic manner how this can be understood from a network circuit model.

\begin{figure}[h]
	\centering
	\includegraphics[width=1.0\columnwidth]{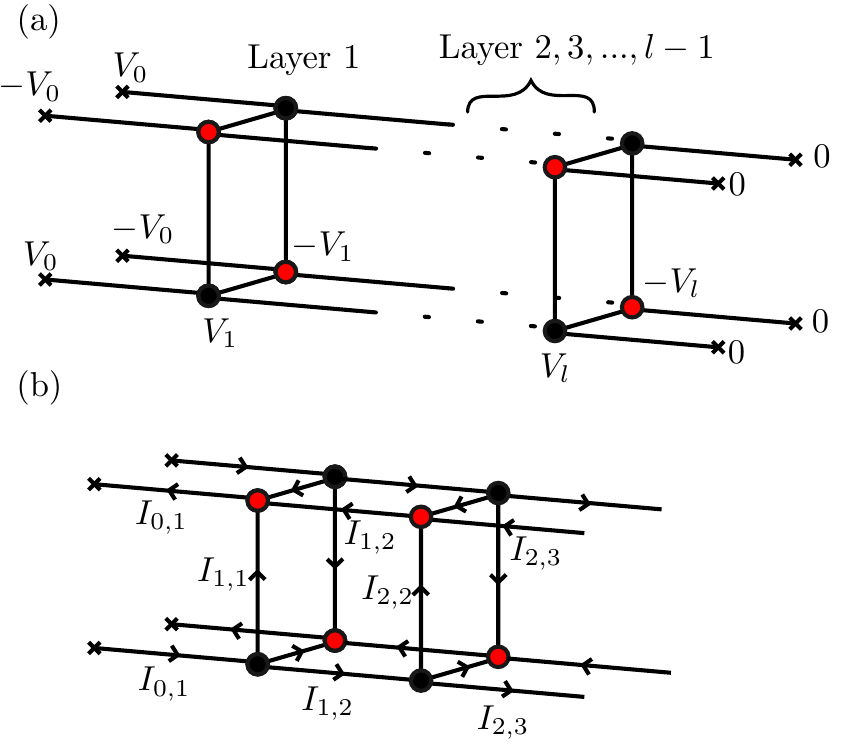}
	\caption{ (a) A section of the network circuit model. There are $l$ transverse layers with associated potentials $V_{i}$ for $i=1,..,l$. Black (red) nodes in layer $i$ have potential $V_{i}$ ($-V_{i}$). To the left, layer 1 is connected to nodes where the potential is staggered: $V_{0}$ or $-V_{0}$. To the right, layer $l$ is connected to nodes with zero potential. Dotted lines indicate where layers $2,3,...,l-1$ are located. The connections between the nodes have the same conductance, shown by solid lines. (b) The leftmost section of the network circuit. The currents and their directions in the circuit are indicated by the arrows. The current between a node in layer $i-1$ and a node in layer $i$ is $I_{i-1,i}$. For the current between two nodes within the same layer $i$ we label the current as $I_{i,i}$. }
	\label{fig:curcuit}
\end{figure}  

We consider a network circuit with $l$ transverse layers of nodes, as shown in Fig.\ \ref{fig:curcuit}. The network is cubic to resemble the tight-binding model. The nodes are connected to adjacent nodes via identical conductances. As a boundary condition, layer 1 is connected to a layer of nodes where the potentials are $V_{0}$ or $-V_{0}$ in a staggered configuration. For simplicity, we use periodic boundary conditions in the two transverse directions. Then, the potentials at the nodes within one transverse layer have the same absolute value but opposite (staggered) signs because of symmetry, in the same way as the leftmost layer in Fig.\ \ref{fig:curcuit}. We associate a potential $V_{1}, V_{2}, ... , V_{l}$ to each layer $1,2, ..., l$, respectively. Layer $l$ is connected to layer $l-1$ on one side and to a layer of nodes with zero potential on the other side. We define currents $I_{i-1,i}$ between adjacent nodes in layer $i-1$ and layer $i$. Similarly, there are currents $I_{i,i}$ between adjacent nodes within the same layer $i$. The directions of these currents are indicated in Fig.\ \ref{fig:curcuit}. From current conservation at each node, we can calculate all potentials and currents in the circuit. 

Kirchoff's law determines the potentials $V_{i}$ for layer $i$ from $l$ linear equations,
\begin{equation}
\begin{pmatrix} 1 & \frac{-1}{10}  & 0 & 0 & ... &   & 0 \\
\frac{-1}{10} & 1 & \frac{-1}{10} & 0 & ... &  & 0 \\
0 & \frac{-1}{10} & 1 & \frac{-1}{10} & ... & & 0 \\
\vdots & & & & & & \vdots \\
0	& ... & &  & \frac{-1}{10} & 1 & \frac{-1}{10} \\
0 & ... & &  & 0 & \frac{-1}{10} & 1
\end{pmatrix}
\begin{pmatrix}
V_{1} \\
V_{2} \\
V_{3} \\
\vdots \\
V_{l\sm1} \\
V_{l}
\end{pmatrix}
= \begin{pmatrix}
V_{0} \\
0 \\
0 \\
\vdots\\
0\\
0
\end{pmatrix}
\label{eq:tridiagonalMatrix} \, ,
\end{equation}
which consists of a uniform, tridiagonal $l\times l$ matrix. 

\begin{figure}[h]
	{	\centering
		\hspace*{-0.5cm}\includegraphics[width=0.45\textwidth]{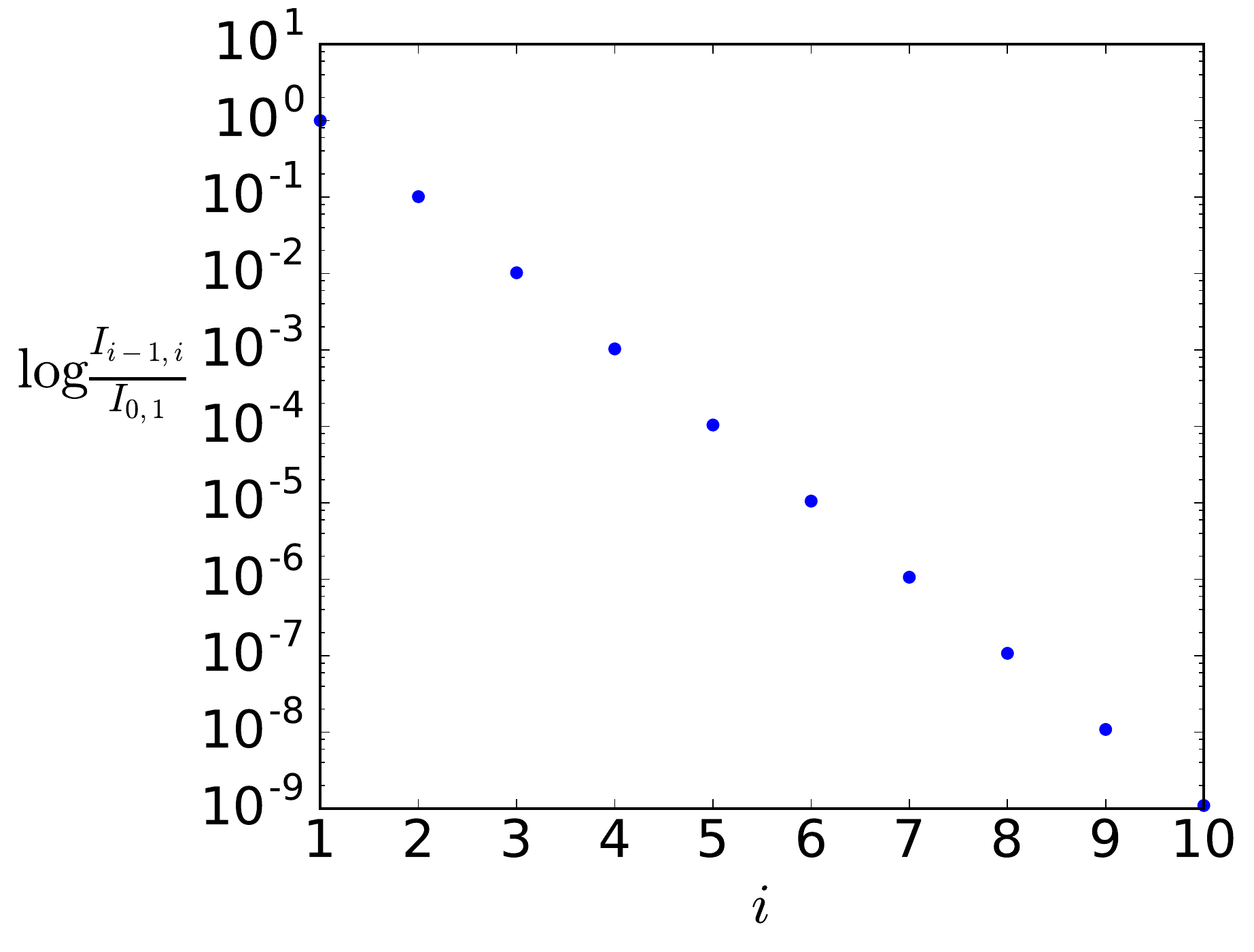}
		\caption{ Logarithmic plot of the current $I_{i-1,i}$ between adjacent nodes in layer $i-1$ and layer $i$. There are $l=10$ transverse layers in the middle of the circuit. The currents are normalized by the current $I_{0,1}$, which is the current between the nodes with potentials $V_{0}$ and $V_{1}$. }
		\label{fig:curcuit_currents_plot}}
\end{figure}

The analytical solution of Eq.\ \eqref{eq:tridiagonalMatrix} for a large integer $l$ is cumbersome. Instead, we present the resulting currents $I_{i-1,i}$ between nodes in layers $i-1$ and $i$ when $l=10$ in Fig.\ \ref{fig:curcuit_currents_plot}. We have verified that the results are similar for a range of different $l$.

The currents $I_{i,i+1}$ between layers $i$ and $i+1$ are reduced by a factor on the order of $10^{-1}$ compared to the currents $I_{i-1,i}$. Similarly, the currents $I_{i,i}$ between nodes in the same layer $i$ are reduced compared to the currents $I_{i-1,i-1}$ in the preceding layer $i-1$. Based on this result, the classical analogue of a staggered current, in the form of Eq.\ \eqref{eq:staggSpinCurrentInLayerX}, decays within few transverse layers.


\section{Numerical results}\label{sec:numerical_results}

We will now consider the effects of spin-conserving and magnetic disorder on the torques. Disorder is modeled using a static random potential $V_{\bs{r}}$. The elastic potential consists of spin-conserving and magnetic-impurity-induced spin-flip scattering parts: 
\begin{equation}
V_{\bs{r}} = V^{\mr{c}}_{\bs{r}}\hat{1}+\bs{V}^{\mr{m}}_{\bs{r}}\cdot\bs{\sigma} \, . 
\end{equation} 
We will consider the effects of spin-conserving and spin-flip scatterings separately. 

The values of $V^{\mr{c}}_{\bs{r}}$ are random in a uniform distribution within the energy range between $-\eta$ and $\eta$. In the case of spin-flip scattering, the magnetic impurities $\bs{V}^{\mr{m}}_{\bs{r}}=V^{\mr{m}}_{\bs{r}}\left(\sqrt{1-\zeta^{2}}\cos\phi,\sqrt{1-\zeta^{2}}\sin\phi,\zeta \right)$, where the parameters are uniformly distributed within $V^{\mr{m}}_{\bs{r}}\in\big[0,\eta_{\mr{m}}\big]$, $\zeta\in\big[-1,1\big]$ and $\phi\in\big[0,2\pi\big)$. In this way, the direction of the magnetic impurities $\bs{V}^{\mr{m}}_{\bs{r}}$ is uniformly distributed on the unit sphere. The disorder then has zero mean, $\left\langle V^{\mr{c}}_{\bs{r}} \right\rangle =\left\langle \bs{V}^{\mr{m}}_{\bs{r}} \right\rangle=0 $, while the variances $\left\langle (V^{\mr{c}}_{\bs{r}})^{2} \right\rangle$ and $\left\langle \left(\bs{V}^{\mr{m}}_{\bs{r}}\right)^{2} \right\rangle$ are given by $\eta^{2}$ and $\eta^{2}_{\mr{m}}$, respectively. The presence of magnetic disorder modifies the spin continuity equations (Eq.\ \eqref{eq:itinerantspincontinuity}) by introducing additional spin-orbit-like terms $\hat{c}^{\dagger}_{\bs{r}}\left(\bs{\sigma}\times\bs{V}^{\mr{m}}_{\bs{r}} \right)\hat{c}_{\bs{r}}$, which we also include in our considerations.

We utilize the python package KWANT \cite{KwantNJP2014} to solve the scattering problem with disorder. In this way, we obtain the wavefunction at all lattice sites. Based on this wavefunction, we compute the spin currents and the torques. The $i$'th component of the torques on the total spin and the spin difference depends on the spin accumulation in the NM:
\begin{subequations}
	\begin{align}
	\tau^{i}_{\mc{M}} =& A^{\mc{M}}_{ij}\mu^{j}_{\mr{S}} \, , \label{eq:torqueMSpinAccResponse} \\
	\tau^{i}_{\mc{N}} =& A^{\mc{N}}_{ij}\mu^{j}_{\mr{S}} \, , \label{eq:torqueNSpinAccResponse}
	\end{align}
\end{subequations}
where summation over repeated indices is implied. The real-valued second-rank tensors $A^{\mc{M}}_{ij}$ and $A^{\mc{N}}_{ij}$ are 
\begin{align}
A^{\mc{M}}_{ij}=& \frac{1}{N_{\mr{AF}}} \frac{\tilde{t}}{4i} \sum_{\bs{r}_{\mperp}} \sum_{n\in\mr{prop}}\sum_{\alpha_{1}\alpha_{2}ss'} \sigma^{ss'}_{i}\sigma^{\alpha_{2}\alpha_{1}}_{j} \nonumber \\ &\times\Big[\psi^{\dagger}_{nE_{\mr{F}}\alpha_{1}}(x=2, \bs{r}_{\mperp},s)\psi_{nE_{\mr{F}}\alpha_{2}}(x=1, \bs{r}_{\mperp},s')  \nonumber \\
&-\psi^{\dagger}_{nE_{\mr{F}}\alpha_{1}}(x=1, \bs{r}_{\mperp},s)\psi_{nE_{\mr{F}}\alpha_{2}}(x=2, \bs{r}_{\mperp},s')\Big] \, ,
\label{ResponseCoeffExpressionMagnetization}
\end{align}
and
\begin{align}
A^{\mc{N}}_{ij}=& \frac{1}{N_{\mr{AF}}} \frac{\tilde{t}}{4i} (\sum_{\bs{r}_{\mperp}\in A} - \sum_{\bs{r}_{\mperp}\in B}) \sum_{n\in\mr{prop}}\sum_{\alpha_{1}\alpha_{2}ss'} \sigma^{ss'}_{i}\sigma^{\alpha_{2}\alpha_{1}}_{j}\nonumber \\ &\times\Big[\psi^{\dagger}_{nE_{\mr{F}}\alpha_{1}}(x=2, \bs{r}_{\mperp},s)\psi_{nE_{\mr{F}}\alpha_{2}}(x=1, \bs{r}_{\mperp},s') - \nonumber \\
&\psi^{\dagger}_{nE_{\mr{F}}\alpha_{1}}(x=1, \bs{r}_{\mperp},s)\psi_{nE_{\mr{F}}\alpha_{2}}(x=2, \bs{r}_{\mperp},s')\nonumber \\
&+2\times\Big(\psi^{\dagger}_{nE_{\mr{F}}\alpha_{1}}(x=1,\bs{r}_{\mperp}+\bs{\delta},s)\psi_{nE_{\mr{F}}\alpha_{2}}(x=1,\bs{r}_{\mperp},s')  \nonumber \\
&- \psi^{\dagger}_{nE_{\mr{F}}\alpha_{1}}(x=1,\bs{r}_{\mperp},s)\psi_{nE_{\mr{F}}\alpha_{2}}(x=1,\bs{r}_{\mperp}+\bs{\delta},s')\Big)\Big] \, ,
\label{ResponseCoeffExpressionNeel}
\end{align}
where the sum is over all propagating modes at the Fermi energy, and we sum terms with $\bs{\delta}=a\hat{y}$ and $\bs{\delta}=a\hat{z}$. For finite systems of length $Na$ in the transverse directions, the end point of the transverse sums with transverse hoppings is $N-1$.

The dominant components of the STTs $\bs{\tau}_{\mc{N}}$ and $\bs{\tau}_{\mc{M}}$ follow from $\bs{\mu}_{\mr{S}}\times\bs{n}$ and $\bs{n}\times(\bs{\mu}_{\mr{S}}\times\bs{n})$, respectively, even for disordered configurations. 

\begin{figure}[h]
	{	\centering
		\hspace*{-0.5cm}\includegraphics[width=0.5\textwidth]{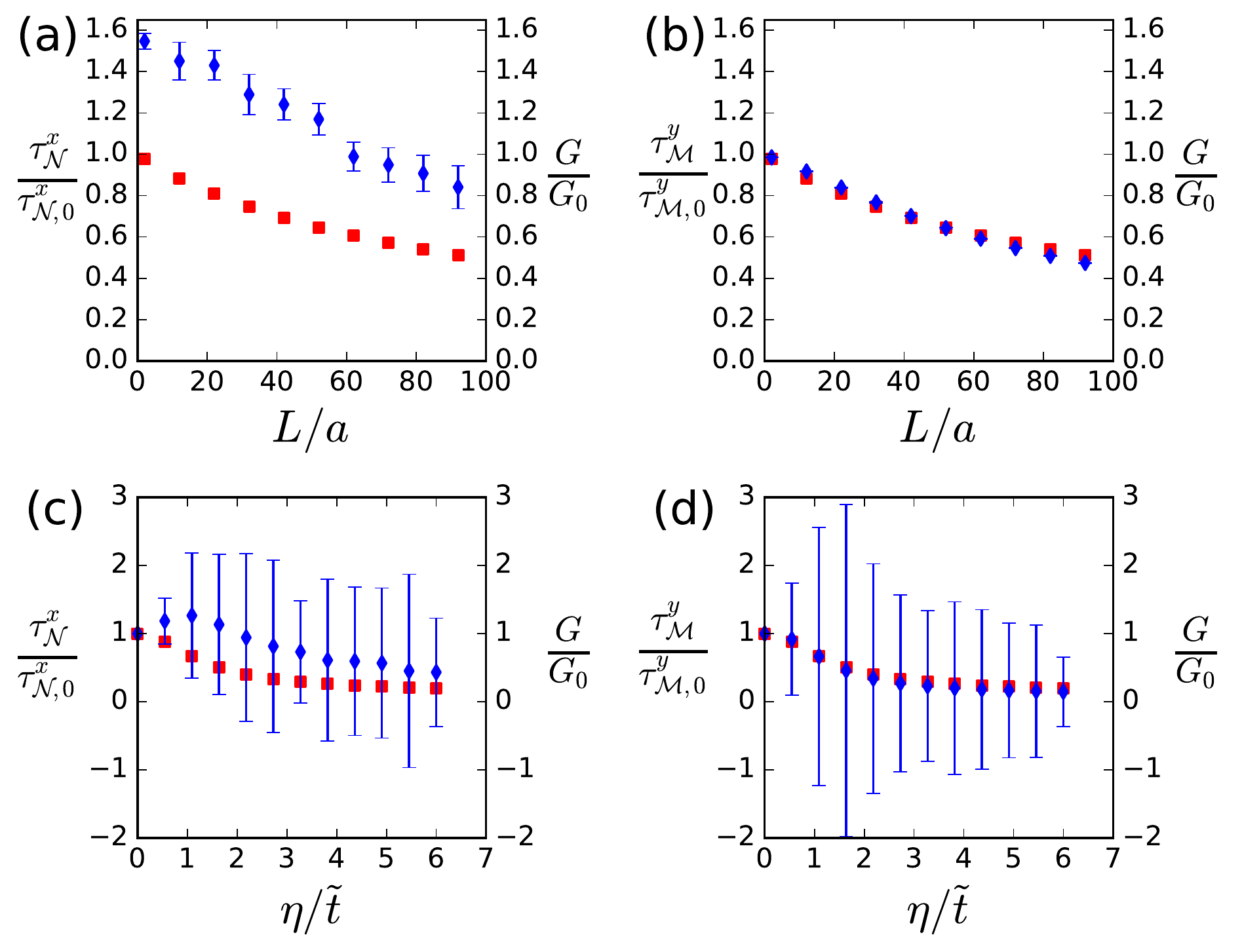}
		\caption{ STTs (blue diamonds) for AFI-NM systems with spin-conserving disorder ($\eta_{\mr{m}}=0$). The Neel order parameter points in $\bs{n}=\hat{z}$, and the torques are induced by a spin accumulation $\bs{\mu}_{\mr{S}}=\mu_{\mr{S}}\hat{y}$. There are $40\times40$ sites in the transverse directions, and the concentration of the impurities is fixed at 0.125. In (a) and (b), we vary the length $L$ of the NM while keeping the disorder-associated potential energy at $\eta =\tilde{t}$. In (c) and (d), we vary the disorder strength $\eta$ for a system with length $L/a = 40$. The red squares show the electrical conductance of a two-terminal system with the same parameters as in the STT cases. The torques are normalized with respect to their corresponding torques, $\tau^{x}_{\mc{N},0}$ and $\tau^{y}_{\mc{M},0}$, when $\eta=0$, i.e., without disorder. $G_{0}$ is the Sharvin conductance. All values are averaged over 15 impurity configurations. The error bars show the standard deviation in the average torques.  }
		\label{fig:STTs_spin-conserving}}
\end{figure}

We show how the relevant STTs are affected by spin-conserving disorder ($\eta_{\mr{m}}=0$) in Fig.\ \ref{fig:STTs_spin-conserving}. In these cases, the AFI has an order parameter $\bs{n}=\hat{z}$, which is transverse to the direction of transport. In the adjacent metal reservoir, a transverse spin accumulation $\bs{\mu}_{\mr{S}}=\mu_{\mr{S}}\hat{y}$ induces the STTs. The exchange coupling was set to $J\mc{S}/\tilde{t} =1$. There are $40\times40$ sites in the two transverse directions. Here, the concentration of the impurities in the NM is constant at 0.125. The STTs are compared to the electrical conductances $G$ of a two-terminal system with the same system parameters as the (one-terminal) systems considered for the STTs, as shown in Fig.\ \ref{fig:STTs_spin-conserving}. $G_{0}$ is the Sharvin conductance. We normalize the STTs by the torques $\tau^{x}_{\mc{N},0}$ and $\tau^{y}_{\mc{M},0}$, which are the torques calculated for systems without disorder. All the values presented in Fig.\ \ref{fig:STTs_spin-conserving} are calculated by averaging over 15 impurity configurations, and the standard deviations are shown as error bars.

In Figs.\ \ref{fig:STTs_spin-conserving}(a) and \ref{fig:STTs_spin-conserving}(b), we consider different lengths $L$ of the disordered region while keeping the impurity-associated parameter $\eta=\tilde{t}$ fixed. In Fig.\ \ref{fig:STTs_spin-conserving}(a), we find that spin-conserving disorder enhances the torque $\tau^{x}_{\mc{N}}$ on $\bs{n}$ compared to the torque without disorder, and the effect is most prominent for disorder close to the interface ($L/a$ less than 62). As the length of the conductor increases, the torque $\tau^{x}_{\mc{N}}/\tau^{x}_{\mc{N},0}$ decreases with a slope similar to the reduction in the conductances $G/G_{0}$. As shown in Fig.\ \ref{fig:STTs_spin-conserving}(b), the normalized values of $\tau^{y}_{\mc{M}}$ follow the conductances very closely as the length of the conductor is changed. 

In Figs.\ \ref{fig:STTs_spin-conserving}(c) and \ref{fig:STTs_spin-conserving}(d), we fix the length $L/a=40$ of the metal and show the torques and conductances for different impurity-associated energies $\eta$. The torque on $\bs{n}$ is enhanced (Fig.\ \ref{fig:STTs_spin-conserving}(c)) for $\eta\sim\tilde{t}$, whereas it decreases for larger $\eta$. For the torque on the magnetization, the mean values $\tau^{y}_{\mc{M}}/\tau^{y}_{\mc{M},0}$ follow the conductance $G/G_{0}$ (Fig.\ \ref{fig:STTs_spin-conserving}(d)). The standard deviations are large in Figs.\ \ref{fig:STTs_spin-conserving}(c) and \ref{fig:STTs_spin-conserving}(d), which indicates that the specific disorder configurations become important when the strength of the impurities may be large.        

\begin{figure}[h]
	{	\centering
		\hspace*{-0.5cm}\includegraphics[width=0.5\textwidth]{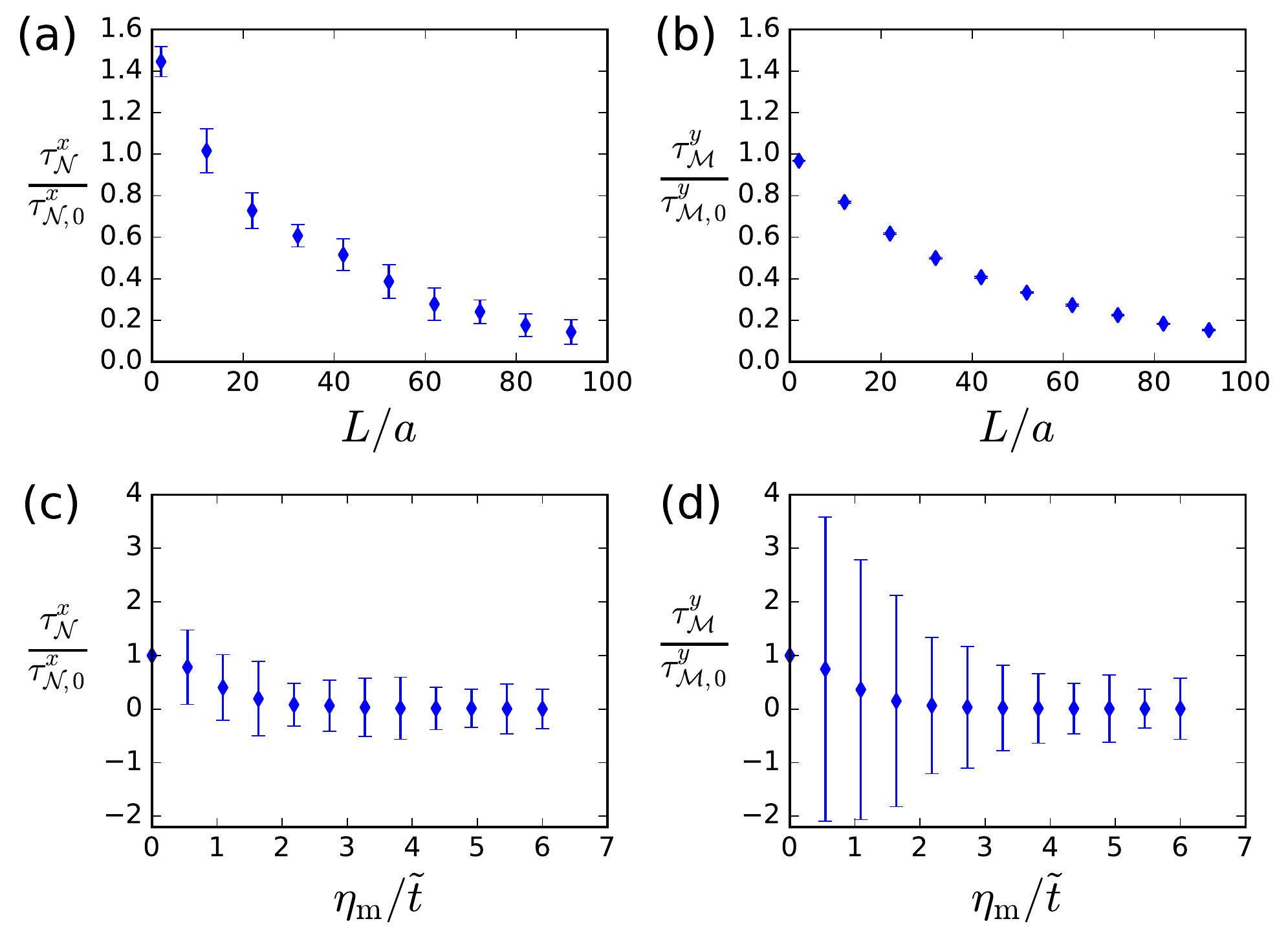}
		\caption{ STTs (blue diamonds) for AFI-NM systems subject to magnetic impurities ($\eta=0$). The static magnetic impurities point in arbitrary directions, with a concentration of 0.125 in the NM. A spin accumulation $\bs{\mu}_{\mr{S}}=\mu_{\mr{S}}\hat{y}$ induces STTs on $\bs{n}=\hat{z}$ and $\bs{m}$. The transverse size is $40\times40$ lattice sites. In (a) and (b), the conductor length $L$ is varied for a fixed value $\eta_{\mr{m}}=\tilde{t}$. In (c) and (d), the length of the disordered region is $L/a=40$, while the parameter $\eta_{\mr{m}}$ varies. The values are normalized with respect to the STTs $\tau^{x}_{\mc{N},0}$ and $\tau^{y}_{\mc{M},0}$ obtained for the systems without disorder. The error bars show the standard deviation of the mean values of the STTs, which are averaged over 15 configurations of the impurities. }
		\label{fig:STT_magnetic_imps}}
\end{figure}

We now consider the STTs on the AFI when all the impurities are magnetic ($\eta=0$), as shown in Fig.\ \ref{fig:STT_magnetic_imps}. Many of the system parameters are the same as the ones considered in the previous paragraphs: Neel order parameter $\bs{n}=\hat{z}$, spin accumulation $\bs{\mu}_{\mr{S}}=\mu_{\mr{S}}\hat{y}$, exchange parameter $J\mc{S}/\tilde{t} =1$, transverse lengths of 40 sites each and an impurity concentration of 0.125. In Figs.\ \ref{fig:STT_magnetic_imps}(a) and \ref{fig:STT_magnetic_imps}(b), we vary the conductor length $L$ when $\eta_{\mr{m}}=\tilde{t}$. The torques  $\tau^{x}_{\mc{N}}$ (Fig.\ \ref{fig:STT_magnetic_imps}(a)) and  $\tau^{y}_{\mc{M}}$ (Fig.\ \ref{fig:STT_magnetic_imps}(b)) considerably decrease as the length of the conductor increases. Spin loss reduces both the spin current and the staggered spin current, resulting in smaller torques on $\bs{m}$ and $\bs{n}$, respectively. In Figs.\ \ref{fig:STT_magnetic_imps}(c) and \ref{fig:STT_magnetic_imps}(d), we change the parameter $\eta_{\mr{m}}$ for a disordered region with length $L/a = 40$. The mean values of the torques $\tau^{x}_{\mc{N}}$ (Fig.\ \ref{fig:STT_magnetic_imps}(c)) and $\tau^{y}_{\mc{M}}$ (Fig.\ \ref{fig:STT_magnetic_imps}(d)) quickly decrease to zero as the parameter $\eta_{\mr{m}}$ increases.  

We have also considered the torques on $\bs{n}$ and $\bs{m}$ for different scattering region sizes and impurity-associated energies when the Neel order parameter points parallel to the transport direction ($\bs{n} = \hat{x}$) and the spin accumulation is transverse (along $\hat{y}$). The torques $-\tau^{z}_{\mc{N}}$ and $\tau^{y}_{\mc{M}}$ (when $\bs{n}=\hat{x}$ and $\bs{\mu}_{\mr{S}}=\mu_{\mr{S}}\hat{y}$) show very similar results compared to the torques $\tau^{x}_{\mc{N}}$ and $\tau^{y}_{\mc{M}}$ (when $\bs{n}=\hat{z}$ and $\bs{\mu}_{\mr{S}}=\mu_{\mr{S}}\hat{y}$), respectively. The STTs are therefore similar for a Neel order parameter that is parallel ($\hat{x}$) or transverse ($\hat{z}$) to the transport direction. This result is reasonable because we average over impurity configurations where the impurity spins point in arbitrary directions.

We also investigated how the spin current $\bs{\mc{I}}^{\mr{s}}_{x}$ and the staggered spin current $\bs{\mathcal{I}}^{\mathrm{ss}}_{x}$ (Eq.\ \eqref{eq:staggSpinCurrentInLayerX}) vary in the transport direction inside the NM. In the presence of magnetic impurities and/or spin-orbit coupling, the spin is not conserved microscopically. The spin current can then be defined in several ways. Our definition (Eq.\ \eqref{eq:spincurrentx}) only includes electron hoppings. As expected, we find that spin-flip-inducing impurities in the NM reduce the spin current (toward the interface) with position $x$. The spin current $\bs{\mc{I}}^{\mr{s}}_{x}$ is unaffected by spin-conserving disorder in the NM. 
\begin{figure}[h]
	{	\centering
		\hspace*{-0.5cm}\includegraphics[width=0.5\textwidth]{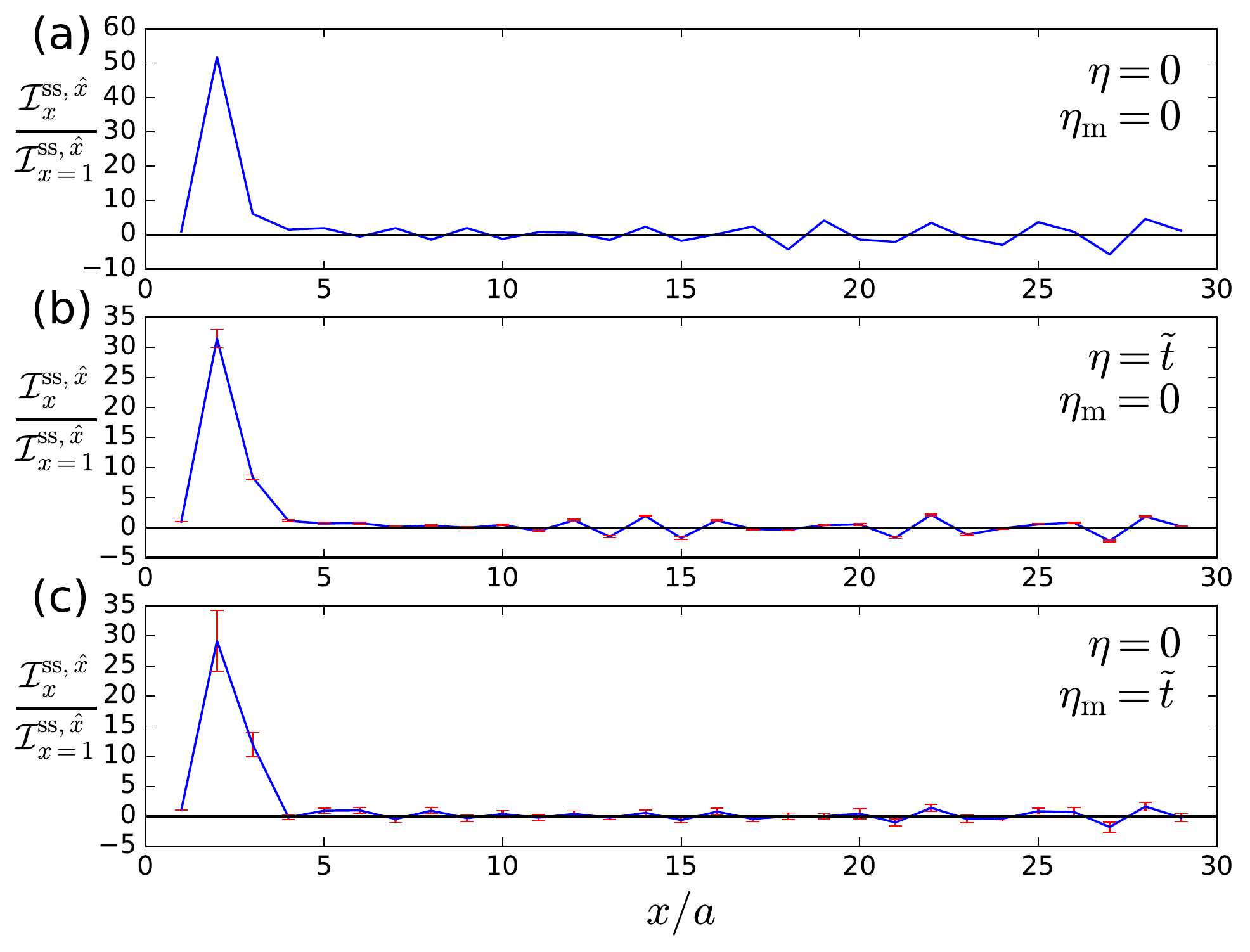}
		\caption{ The staggered spin current $\mathcal{I}^{\mr{ss},\hat{x}}_{x}$ as a function of position $x$ inside the NM from the left (near the interface) to the right (in the NM bulk). The length of the NM is $L=30a$, and there are 40 sites in the two transverse directions. The position $x/a=1$ is directly at the AFI-NM interface. We separate three cases: (a) no disorder, (b) only spin-conserving disorder ($\eta=\tilde{t}$, $\eta_{\mr{m}}=0$) and (c) only magnetic disorder ($\eta=0$, $\eta_{\mr{m}}=\tilde{t}$). The staggered spin currents are normalized to their value at the interface, $\mathcal{I}^{\mr{ss},\hat{x}}_{x=1}$. In (b) and (c), the concentrations of the impurities are 0.125. A spin accumulation $\bs{\mu}_{\mr{S}}=\mu_{\mr{S}}\hat{y}$ in the reservoir is assumed.  }
		\label{fig:staggered_spin_currents_kwant}}
\end{figure}

In the network circuit model, a staggered current vanishes within few lattice constants, as shown in Fig.\ \ref{fig:curcuit_currents_plot} from section\ \ref{section:decay_stagg_current}. However, in the coherent and finite-size regime, the staggered spin current $\bs{\mathcal{I}}^{\mathrm{ss}}_{x}=\bs{\mathcal{I}}^{\mathrm{s},A}_{x}-\bs{\mathcal{I}}^{\mathrm{s},B}_{x}+2\bs{\mathcal{I}}^{AB}_{x}$ inside the NM bulk exhibits a slightly different behavior. In Fig.\ \ref{fig:staggered_spin_currents_kwant}, the staggered spin current $\bs{\mathcal{I}}^{\mathrm{ss}}_{x}$ obtains irregular values around 0, on the order of plus/minus $\bs{\mathcal{I}}^{\mathrm{ss}}_{x=1}$, which is the value measured at the interface. At $x=2$, there is also a peak in $\bs{\mathcal{I}}^{\mathrm{ss}}_{x}$, which is generally much larger than the staggered spin current directly at the interface, $x=1$, even with or without disorder. The system considered in Fig.\ \ref{fig:staggered_spin_currents_kwant} has length $L/a = 30$, $\bs{n}=\hat{z}$, $\bs{\mu}_{\mr{S}}=\mu_{\mr{S}}\hat{y}$ and 40 sites in the two transverse directions. Here, we separated three cases where there is no disorder (Fig.\ \ref{fig:staggered_spin_currents_kwant}(a)), spin-conserving disorder (Fig.\ \ref{fig:staggered_spin_currents_kwant}(b)) and magnetic disorder (Fig.\ \ref{fig:staggered_spin_currents_kwant}(c)). Even for the cases without disorder, there are non-zero irregular and fluctuating values for the staggered spin current.


\section{Conclusions}\label{sec:conclusions}

The microscopic spin currents at the AFI-metal interface determine the STTs. The torques on the magnetization $\bs{m}$ and the Neel order parameter $\bs{n}$ are proportional to the spin current $\bs{\mc{I}}^{\mr{s}}_{1}$ and the staggered spin current $\bs{\mc{I}}^{\mr{ss}}_{1}$, respectively. In a reciprocal way, a precessing AF Neel order parameter pumps both a spin current and a staggered spin current. In a manner similar to spin pumping in ferromagnets, one can calculate the AF-pumped spin currents by transforming to a rotating spin frame of reference. 

The STTs on the AFI are affected by spin-conserving and/or spin-flip-inducing disorder in the NM. These effects are particularly prominent when the disorder is very close to the AFI-NM interface. Spin-conserving disorder reduces the torques on $\bs{n}$ and $\bs{m}$, in a manner similar to Ohm's law. Magnetic impurities result in spin loss and reduced torques. However, for intermediate-strength non-magnetic impurities close to the interface, we find that the torque on the Neel order parameter is enhanced compared with the corresponding torque without impurities. 

For a finite tight-binding system, the staggered spin current $\bs{\mc{I}}^{\mr{ss}}_{x}$ rapidly becomes small and irregular away from the AFI-NM interface, both with or without disorder. The staggered spin current is not a conserved quantity, even when the spin is conserved.  This behavior is well captured by a classical circuit model, where the staggered current quickly decays away from an interface with staggered potentials. 

The research leading to these results has received funding from the European Research Council via Advanced Grant number 669442 "Insulatronics" as well as the Research Council of Norway via Grant no. 239926 and through its Centres of Excellence
"QuSpin".


\appendix

\section{Rate of change of the spins}
The rate of change of an operator $\hat{\mc{O}}(t) = e^{i\hat{H}t/\hbar}\hat{\mc{O}}e^{-i\hat{H}t/\hbar}$ is calculated via its commutator with the Hamiltonian, i.e., $\hbar\partial_{ t}\hat{\mc{O}}(t)=i\big[\hat{H},\hat{\mc{O}}(t) \big]$.
We use that the localized spins in the magnetic insulator obey $\big[\mathcal{\hat{S}}^{(j)}_{\bs{r}},\mathcal{\hat{S}}^{(j')}_{\bs{r}'}\big]=\delta_{\bs{r}',\bs{r}}i\hbar\epsilon_{jj'j''}\mathcal{\hat{S}}^{(j'')}_{\bs{r}}$, where summation over spatial directions, $j=x, y$ or $z$ for $j, j', j''$, is implied ($\epsilon_{jj'j''}$ is the Levi-Civita tensor). Then,
the rate of change of the localized AFI spins at the interface is
\begin{align}
\frac{\partial\bs{\mathcal{\hat{S}}}_{0,\bs{r}_{\perp}}(t)}{\partial t}
=& J\bs{\mathfrak{\hat{s}}}_{1,\bs{r}_{\perp}}(t)\times\bs{\mathcal{\hat{S}}}_{0,\bs{r}_{\perp}}(t) \, ,
\label{eq:localizedSpinsRateOfChange}
\end{align}
in terms of the itinerant spins $\bs{\mathfrak{\hat{s}}}_{1,\bs{r}_{\perp}}$ at the interface (in the layer at $x=1$).

To find spin continuity equations for the itinerant spins, we write the kinetic part $\hat{H}_{\mr{hop}}$ of the Hamiltonian in Eq.\ \eqref{eq:HtotNM}, with the notation $\bs{r}=(x,y,z)$, as
\begin{align}
\hat{H}_{\mr{hop}}=&-\tilde{t}\sum_{x=1}^{\infty}\sum_{y=1}^{N_{y}}\sum_{z=1}^{N_{z}}\big(\hat{c}^{\dagger}_{\bs{r}\smm\bs{\delta}_{x}}\hat{c}_{\bs{r}}+\hat{c}^{\dagger}_{\bs{r}\smm\bs{\delta}_{y}}\hat{c}_{\bs{r}}(1-\delta_{y,N_{y}})\nonumber \\
&+\hat{c}^{\dagger}_{\bs{r}\smm\bs{\delta}_{z}}\hat{c}_{\bs{r}}(1-\delta_{z,N_{z}})+\mathrm{h.c.}\big) \, ,
\end{align}
where h.c. is the Hermitian conjugate. Expressions for the spin currents are found by commuting $\hat{H}_{\mr{hop}}$ with the itinerant spin density. The electron operators obey the commutator $\left[\hat{c}^{\dagger}_{\bs{r}_{1}s_{1}}\hat{c}_{\bs{r}_{2}s_{2}}, \hat{c}^{\dagger}_{\bs{r}s}\hat{c}_{\bs{r}s'} \right]= \hat{c}^{\dagger}_{\bs{r}_{1}s_{1}}\hat{c}_{\bs{r}s'}\delta_{\bs{r}_{2},\bs{r}}\delta_{s_{2},s}-\hat{c}^{\dagger}_{\bs{r}s}\hat{c}_{\bs{r}_{2}s_{2}}\delta_{\bs{r}_{1},\bs{r}}\delta_{s_{1},s'}$, when writing the spin indices ($s$ and so on) explicitly. As an example, the rate of change of spin density in the $\hat{x}$-direction follows from
\begin{align}
&\sum_{x'=1}^{\infty}\sum_{y'=1}^{N_{y}}\sum_{z'=1}^{N_{z}}\left[\hat{c}^{\dagger}_{\bs{r}'\smm\bs{\delta}_{x}}\hat{c}_{\bs{r}'}+\hat{c}^{\dagger}_{\bs{r}'}\hat{c}_{\bs{r}'\smm\bs{\delta}_{x}},\hat{c}^{\dagger}_{\bs{r}}\bs{\sigma}\hat{c}_{\bs{r}} \right]  \nonumber \\
=&\hat{c}^{\dagger}_{\bs{r}\smm\bs{\delta}_{x}}\bs{\sigma}\hat{c}_{\bs{r}}-\hat{c}^{\dagger}_{\bs{r}}\bs{\sigma}\hat{c}_{\bs{r}\smm\bs{\delta}_{x}} \nonumber \\
&+(1-\delta_{1,x})\big(\hat{c}^{\dagger}_{\bs{r}\sm\bs{\delta}_{x}}\bs{\sigma}\hat{c}_{\bs{r}}-\hat{c}^{\dagger}_{\bs{r}}\bs{\sigma}\hat{c}_{\bs{r}\sm\bs{\delta}_{x}} \big) \, ,
\end{align}
which via the spin continuity equation (Eq.\ \eqref{eq:itinerantspincontinuity}) yields our definition of the spin current in Eq.\ \eqref{eq:spincurrentx}. Similarly, we find $\bs{\hat{j}}^{\pmm\hat{y}}_{\bs{r}}$ and $\bs{\hat{j}}^{\pmm\hat{z}}_{\bs{r}}$.

We also use that $\left[\sigma_{j},\sigma_{j'}\right]=2i\varepsilon_{jj'j''}\sigma_{j''}$ to calculate
\begin{align}
\frac{i}{\hbar}\left[\hat{H}_{\mr{AF,N}}, \frac{\hbar}{2}\hat{c}^{\dagger}_{\bs{r}}\bs{\sigma}\hat{c}_{\bs{r}} \right]
=&\delta_{1,x}J\bs{\mc{\hat{S}}}_{0,\bs{r}_{\mperp}}\times\bs{\mf{\hat{s}}}_{1,\bs{r}_{\mperp}} \, ,
\label{eq:commutator_itspin_Haf}
\end{align}
which relates the rate of change of the localized AFI spins to the itinerant spins in the NM (Eqs.\ \eqref{eq:localizedSpinsRateOfChange} and \eqref{eq:commutator_itspin_Haf}).

In total, the spin continuity equation is
\begin{align}
0 =&\frac{\partial\hat{\bs{\mf{s}}}_{\bs{r}}}{\partial t}+
\bs{\hat{j}}^{\smm\hat{x}}_{\bs{r}}+(1-\delta_{1,x})\bs{\hat{j}}^{\sm\hat{x}}_{\bs{r}}+(1-\delta_{N_{y},y})\bs{\hat{j}}^{\smm\hat{y}}_{\bs{r}} \nonumber \\
&+ (1-\delta_{1,y})\bs{\hat{j}}^{\sm\hat{y}}_{\bs{r}}+(1-\delta_{N_{z},z})\bs{\hat{j}}^{\smm\hat{z}}_{\bs{r}}+(1-\delta_{1,z})\bs{\hat{j}}^{\sm\hat{z}}_{\bs{r}}\nonumber \\ 
&+\hat{c}^{\dagger}_{\bs{r}}\left(\bs{\sigma}\times\bs{V}^{\mr{m}}_{\bs{r}} \right)\hat{c}_{\bs{r}}+\delta_{1,x}J\bs{\mf{\hat{s}}}_{1,\bs{r}_{\mperp}}\times\bs{\mc{\hat{S}}}_{0,\bs{r}_{\mperp}} \, ,
\label{eq:itinerantspincontinuity}
\end{align}
where the terms $\hat{c}^{\dagger}_{\bs{r}}\left(\bs{\sigma}\times\bs{V}^{\mr{m}}_{\bs{r}} \right)\hat{c}_{\bs{r}}$ arise from magnetic impurities. 

\section{Expectation values}\label{appendix:expectation_values}
To calculate the STTs, we evaluate spin currents between lattice sites, similar to the expression in Eq.\ \eqref{eq:spincurrentx}, in all three spatial directions. In general, we need to calculate the expectation value of the quantity
$\bs{J}(\bs{r},\bs{\delta}) = \left\langle \frac{\tilde{t}}{2i}\left(\hat{c}^{\dagger}_{\bs{r}+\bs{\delta}}\bs{\sigma}\hat{c}_{\bs{r}}-\hat{c}^{\dagger}_{\bs{r}}\bs{\sigma}\hat{c}_{\bs{r}+\bs{\delta}} \right)  \right\rangle $, where $\bs{\delta}=\{\pm\hat{x}, \pm\hat{y}, \pm\hat{z}\}$. The amplitude at a node inside the scattering region depends on the scattering states incoming from the NM reservoir. In this way, the expectation value of quantities inside the scattering region can be found via the distribution function in the reservoir $f_{\alpha\beta}(E)$. For a static scattering problem, the field operator inside the scattering region can be expanded in wavefunctions similar to Eq.\ \eqref{eq:DefinitionOfTheFieldOperator}. For the disordered cases, we find all $\psi_{nE\alpha}(\bs{r})$ inside the scattering region by using the python package KWANT \cite{KwantNJP2014}, which solves the scattering problem by matching of the wave functions.

In linear response, one can write the part of the distribution function (from Eq.\ \eqref{eq:distributionFunction}) that describes the spin accumulation as 
\begin{align}
\bs{f}_{\mr{S}}(E) =& \frac{\bs{\mu}_{\mr{S}}}{2|\bs{\mu}_{\mr{S}}|}\left(f_{\mr{FD}}(E-\mu_{\uparrow}) - f_{\mr{FD}}(E-\mu_{\downarrow}) \right) \, ,
\label{eq:spinAccumulationDeltaFunctionApproximation}
\end{align}
by defining chemical potentials $\mu_{\uparrow}$ ($\mu_{\downarrow}$) for spin up (down) along the direction of the spin accumulation $|\bs{\mu_{\mr{S}}}| = \mu_{\uparrow}-\mu_{\downarrow}$. By Taylor expansion of the FD distributions to first order around the equilibrium chemical potential and by using $-\partial f_{\mr{FD}}/\partial E|_{\mu=E_{\mr{F}}} \approx \delta(E-E_{\mr{F}})$ in the low-temperature regime, then $\bs{f}_{\mr{S}}(E) \approx (1/2)\bs{\mu}_{\mr{S}}\delta(E-E_{\mr{F}})$, which we use in the following. 

We now write the spin projections $s$ explicitly with the notation $\psi_{nE\alpha}(\bs{r},s)$ and use Eqs.\ \eqref{eq:DefinitionOfTheFieldOperator}, \eqref{eq:expectationValueReservoir} and \eqref{eq:distributionFunction} to express the quantity $\bs{J}(\bs{r},\bs{\delta})$ as
\begin{align}
\bs{J}(\bs{r},\bs{\delta}) =& \frac{\tilde{t}}{2i} \sum_{n=1}^{N_{\mperp}}\sum_{\alpha_{1}\alpha_{2}ss'}\int_{\varepsilon^{\mperp}_{n}}^{\infty} dE~\bs{\sigma}_{ss'}f_{\alpha_{2}\alpha_{1}}(E) \nonumber \\
&\Big(\psi^{\dagger}_{nE\alpha_{1}}(\bs{r}+\bs{\delta},s)\psi_{nE\alpha_{2}}(\bs{r},s')  \nonumber \\
&- \psi^{\dagger}_{nE\alpha_{1}}(\bs{r},s)\psi_{nE\alpha_{2}}(\bs{r}+\bs{\delta},s')\Big) \, ,
\end{align}
which we separate in two terms: $\bs{J}(\bs{r},\bs{\delta}) \equiv \bs{J}_{1}(\bs{r},\bs{\delta}) + \bs{J}_{2}(\bs{r},\bs{\delta})$. The term $\bs{J}_{1}(\bs{r},\bs{\delta})$ is independent of the spin accumulation and can be evaluated in the low-temperature limit by approximating $f_{\mr{FD}}(E-\mu) \approx \Theta(E_{\mr{F}}-E)$ as a Heaviside step function:
\begin{align}
\bs{J}_{1}(\bs{r},\bs{\delta}) =&\frac{\tilde{t}}{2i} \sum_{n=1}^{N_{\mperp}}\sum_{\alpha ss'}\int_{\varepsilon^{\mperp}_{n}}^{E_{\mr{F}}} dE \Big(\psi^{\dagger}_{nE\alpha}(\bs{r}+\bs{\delta},s)\psi_{nE\alpha}(\bs{r},s')  \nonumber \\
&- \psi^{\dagger}_{nE\alpha}(\bs{r},s)\psi_{nE\alpha}(\bs{r}+\bs{\delta},s')\Big)\bs{\sigma}^{ss'} \, .
\end{align}
The spin-accumulation-induced term $\bs{J}_{2}(\bs{r},\bs{\delta})$ is calculated with the distribution from Eq.\ \eqref{eq:spinAccumulationDeltaFunctionApproximation} in linear response, which yields 
\begin{align}
\bs{J}_{2}(\bs{r},\bs{\delta}) =& \frac{\tilde{t}}{4i} \sum_{n\in\mr{prop}}\sum_{\alpha_{1}\alpha_{2}ss'}\bs{\sigma}^{ss'}(\bs{\sigma}^{\alpha_{2}\alpha_{1}}\cdot\bs{\mu}_{\mr{S}}) \nonumber \\
&\times\Big(\psi^{\dagger}_{nE_{\mr{F}}\alpha_{1}}(\bs{r}+\bs{\delta},s)\psi_{nE_{\mr{F}}\alpha_{2}}(\bs{r},s')  \nonumber \\
&-\psi^{\dagger}_{nE_{\mr{F}}\alpha_{1}}(\bs{r},s)\psi_{nE_{\mr{F}}\alpha_{2}}(\bs{r}+\bs{\delta},s')\Big) \, ,
\end{align}
where the sum is over the modes that are propagating at the Fermi energy.
The torques on the total spin and the spin difference in the AFI are then found by summing currents in the form of $\bs{J}(\bs{r},\bs{\delta})$ in terms of the total spin current and staggered spin current at the interface, respectively. From Eqs. \eqref{eq:torqueMagnetizationAsSpinCurrent} and \eqref{eq:torqueNeelAsStaggSpinCurrent},
\begin{subequations}
	\begin{align}
	\bs{\tau}_{\mc{M}} =& \frac{1}{N_{\mathrm{AF}}} \sum_{\bs{r}_{\mperp}}\bs{J}(x=1, \bs{r}_{\mperp}, \bs{\delta} = \hat{x}) \, , \label{eq:torqueMfromJ} \\
	\bs{\tau}_{\mc{N}} =& \frac{1}{N_{\mathrm{AF}}} (\sum_{\bs{r}_{\mperp}\in A} - \sum_{\bs{r}_{\mperp}\in B})\Big(\bs{J}(x=1, \bs{r}_{\mperp}, \bs{\delta} = \hat{x}) \nonumber \\
	&+2\left(\bs{J}(x=1, \bs{r}_{\mperp}, \bs{\delta} = \hat{y}) + \bs{J}(x=1, \bs{r}_{\mperp}, \bs{\delta} = \hat{z})\right) \Big) \, , \label{eq:torqueNfromJ}
	\end{align}
\end{subequations}
where the transverse sums with transverse hoppings stop at $N-1$ for finite lengths $Na$ in the transverse directions. The response coefficients $A^{\mc{M}}_{ij}$ and $A^{\mc{N}}_{ij}$ in Eqs.\ \eqref{eq:torqueMSpinAccResponse} and \eqref{eq:torqueNSpinAccResponse} then follow by evaluating $\bs{J}_{2}(\bs{r},\bs{\delta})$ in Eqs. \eqref{eq:torqueMfromJ} and \eqref{eq:torqueNfromJ}.

We have numerically verified for some relevant cases that the contributions from $\bs{J}_{1}(\bs{r},\bs{\delta})$ to the torques vanish when the disorder is spin conserving. However, for some special configurations with magnetic impurities, there may be small but non-zero torques, even in equilibrium ($\bs{\mu}_{\mr{S}}=0$). Such equilibrium torques represent an additional anisotropy in the magnetic system. 

\section{Gauge transformation}\label{appendix:gauge_transformation}
The gauge transformation in Sec.\ \ref{section:theory} is explained in detail in the following. When the AFI spins precess, the evolution of the electron operators $\hat{c}_{\bs{r}}(t)$ are governed by the Hamiltonian $\hat{H}(t) = \hat{H}_{\mr{NM}} + \hat{H}_{\mr{AF,N}}(t)$, where these two terms are given by Eqs.\ \eqref{eq:HtotNM} and \eqref{eq:hamiltonian_dynamic_exchange}, respectively. To find the spin currents pumped at the AFI-NM interface, we consider that there are no {\it magnetic} impurities in $\hat{H}_{\mr{NM}}$. The classical AFI spin vectors at the interface, $\bs{\mc{S}}_{0,\bs{r}_{\mperp}}(t) = \pm\mc{S}\bs{n}(t)$, with length $\mc{S}$, are parametrized by the Neel order parameter $\bs{n}(t) = (\sin\theta_{0}\cos\omega_{0} t,\sin\theta_{0}\sin\omega_{0} t,\cos\theta_{0})$ in terms of precession frequency $\omega_{0}$ and the angle $\theta_{0}$ with respect to the $\hat{z}$-axis. The sign + (-) of $\bs{\mc{S}}_{0,\bs{r}_{\mperp}}(t)$ is determined by which sublattice $\mc{A}$ ($\mc{B}$) the spins reside on. In the frame where the spin quantization axis rotates together with $\bs{n}(t)$, the evolution of $\hat{\tilde{c}}_{\bs{r}}(t) = U(t)\hat{c}_{\bs{r}}(t)$ is described by the Hamiltonian $\hat{\tilde{H}}$. We denote Hamiltonians in the first quantization without a hat ($~\hat{}~$), i.e., $\hat{H} = \sum_{\bs{r}\bs{r}'}\hat{c}^{\dagger}_{\bs{r}}H_{\bs{r},\bs{r}'}\hat{c}_{\bs{r}'}$, and similarly, $\hat{\tilde{H}} = \sum_{\bs{r}\bs{r}'}\hat{\tilde{c}}^{\dagger}_{\bs{r}}\tilde{H}_{\bs{r},\bs{r}'}\hat{\tilde{c}}_{\bs{r}'}$.

In a Lagrangian approach, the time evolution of the operators $\hat{c}_{\bs{r}}(t)$ is described by the Lagrangian
\begin{equation}
\mc{L} = i\hbar\sum_{\bs{r}}\hat{c}^{\dagger}_{\bs{r}}(t)\partial_{t}\hat{c}_{\bs{r}}(t) -\sum_{\bs{r}\bs{r}'}\hat{c}^{\dagger}_{\bs{r}}(t)H_{\bs{r},\bs{r}'}\hat{c}_{\bs{r}}(t)\, ,
\label{eq:Lagrangian}
\end{equation}
and the Euler-Lagrange equation
\begin{equation}
\frac{d}{dt}\frac{\partial\mc{L}}{\partial(\partial_{t}\hat{c}^{\dagger}_{\bs{r}})} = \frac{\partial\mc{L}}{\partial \hat{c}^{\dagger}_{\bs{r}}} \, . \label{eq:Euler-Lagrange}
\end{equation}
The evolution of the operators $\hat{\tilde{c}}_{\bs{r}}(t)$ is described in a similar way by changing $\hat{c}_{\bs{r}}\rightarrow\hat{\tilde{c}}_{\bs{r}}$ in Eq.\ \eqref{eq:Euler-Lagrange}. We use the unitarity of $U(t)$ and insert $\hat{c}_{\bs{r}}(t) = U^{\dagger}(t)\hat{\tilde{c}}_{\bs{r}}(t)$ and its Hermitian conjugate into the Lagrangian in Eq.\ \eqref{eq:Lagrangian}, which yields
\begin{align}
\mc{L} =& i\hbar\sum_{\bs{r}}\hat{\tilde{c}}^{\dagger}_{\bs{r}}(t)UU^{\dagger}\partial_{t}\hat{\tilde{c}}_{\bs{r}}(t) + i\hbar\sum_{\bs{r}}\hat{\tilde{c}}^{\dagger}_{\bs{r}}(t)U\frac{\partial U^{\dagger}}{\partial t}\hat{\tilde{c}}_{\bs{r}}(t) \nonumber \\
&-\sum_{\bs{r}\bs{r}'}\hat{c}^{\dagger}_{\bs{r}}(t)UH_{\bs{r},\bs{r}'}U^{\dagger}\hat{c}_{\bs{r}'}(t) \nonumber \\
= & i\hbar\sum_{\bs{r}}\hat{\tilde{c}}^{\dagger}_{\bs{r}}(t)\partial_{t}\hat{\tilde{c}}_{\bs{r}}(t) - \sum_{\bs{r}\bs{r}'}\hat{\tilde{c}}^{\dagger}_{\bs{r}}(t)\tilde{H}_{\bs{r},\bs{r}'}\hat{\tilde{c}}_{\bs{r}'}(t) \, ,
\end{align}
in terms of 
\begin{equation}
\tilde{H}_{\bs{r},\bs{r}'} = UH_{\bs{r},\bs{r}'}U^{\dagger} - i\hbar U\frac{\partial U^{\dagger}}{\partial t}\delta_{\bs{r}\bs{r}'} \, .
\end{equation}
The Hamiltonian $\tilde{H}$ then includes the gauge potential $-i\hbar U\frac{\partial U^{\dagger}}{\partial t}$, which is present at all sites.  

\section{Umklapp scattering}

To calculate the torques in Eq.\ \eqref{eq:the_torques}, we used periodic boundary conditions in the two transverse directions. Then, the longitudinal momentum $k^{\parallel}_{n_{y}n_{z}}(E)$ of the propagating states is determined via the energy dispersion $\cos k^{\parallel}_{n_{y}n_{z}}a = 3-E/(2\tilde{t}) -\cos(\frac{2\pi n_{y}}{N}) -\cos(\frac{2\pi n_{z}}{N})$.

We solve the scattering problem via wavefunction matching with the ansatz that the scattering coefficients include both normal and umklapp scattering:  $\frac{1}{2}(r^{\uparrow}_{m_{y}m_{z},n_{y}n_{z}}+r^{\downarrow}_{m_{y}m_{z},n_{y}n_{z}}) =A_{n_{y}n_{z}}\delta_{m_{y},n_{y}}\delta_{m_{z},n_{z}}$ and $\frac{1}{2}(r^{\uparrow}_{m_{y}m_{z},n_{y}n_{z}}-r^{\downarrow}_{m_{y}m_{z},n_{y}n_{z}}) = B_{n_{y}n_{z}}\delta_{m_{y},\bar{n}_{y}}\delta_{m_{z},\bar{n}_{z}}$. In terms of the parameter $\lambda =J\mc{S}/\tilde{t}$, the normal and umklapp scattering reflection coefficients are
\begin{subequations}\label{eq:scatteringCoeffs}
	\begin{align}
	A_{n_{y}n_{z}} (E) =& \frac{-1+\lambda^{2}e^{i(k^{\parallel}_{\bar{n}_{y}\bar{n}_{z}}-k^{\parallel}_{n_{y}n_{z}})a}}{1-\lambda^{2}e^{i(k^{\parallel}_{\bar{n}_{y}\bar{n}_{z}}+k^{\parallel}_{n_{y}n_{z}})a}} \, ,
	\label{eq:normalScattCoeff} \\ 
	B_{n_{y}n_{z}} (E) =& \lambda \sqrt{\frac{v_{\bar{n}_{y}\bar{n}_{z}}}{v_{n_{y}n_{z}}}} \frac{2i\sin k^{\parallel}_{n_{y}n_{z}}a}{1-\lambda^{2}e^{i(k^{\parallel}_{\bar{n}_{y}\bar{n}_{z}}+k^{\parallel}_{n_{y}n_{z}})a}} \, ,
	\label{eq:umklappScattCoeff}
	\end{align}
\end{subequations}
respectively, with quantum numbers $\bar{n}_{y}$ and $\bar{n}_{z}$ as defined in the main text.

The wavefunction within the scattering region is 
\begin{align}
\psi_{n_{y}n_{z}E\alpha}(x=1,\bs{r}_{\mperp},s) =& \frac{1}{\sqrt{hv_{n_{y}n_{z}}}} \frac{(-2i)\sin k^{\parallel}_{n_{y}n_{z}}a }{1-\lambda^{2}e^{i(k^{\parallel}_{\bar{n}_{y}\bar{n}_{z}}+k^{\parallel}_{n_{y}n_{z}})a}} \nonumber \\
& \times\Big[ \delta_{\alpha,s} \varphi_{n_{y}n_{z}}(\bs{r}_{\mperp}) \nonumber \\
&- \lambda (\bs{n}\cdot\bs{\sigma}^{s\alpha}) \varphi_{\bar{n}_{y}\bar{n}_{z}}(\bs{r}_{\mperp}) \Big] \, . \label{eq:scatteringWfCleanCase}
\end{align}
These solutions are such that in the limits $JS/\tilde{t}\rightarrow 0$ and $JS/\tilde{t}\rightarrow \pm\infty$, the scattering problem is reduced to hard wall scattering at $x=1$ and $x=2$, respectively. Expressions for the spin currents are found by evaluating $A_{n_{y}n_{z}}(E)$ and $B_{n_{y}n_{z}}(E)$ at the Fermi energy $E_{\mr{F}}=6\tilde{t}$ (half-filling). In this case, the umklapp-scattered longitudinal momentum is $k^{\parallel}_{\bar{n}_{y}\bar{n}_{z}}(E_{\mr{F}}) = \pi/a - k^{\parallel}_{n_{y}n_{z}}(E_{\mr{F}})$ and $v_{\bar{n}_{y}\bar{n}_{z}}(E_{\mr{F}}) = v_{n_{y}n_{z}}(E_{\mr{F}})$. The staggered spin current is calculated by using the wavefunctions in Eq.\ \eqref{eq:scatteringWfCleanCase} directly. The spin current can be calculated either from the wavefunctions at the interface or as the real part of $g^{\uparrow\downarrow} = N_{\mr{p}}-\sum_{mn}r^{\uparrow}_{mn}(r^{\downarrow}_{mn})^{*}$, where the transverse quantum labels are $n\rightarrow (n_{y},n_{z})$ such that $\sum_{n}\rightarrow\sum_{n_{y}n_{z}}$ and similarly for $m$.



%

\end{document}